%% file: Zn_cathode_paper.tex
\documentclass[
journal = jpclcd,
manuscript = article
]{achemso}

\usepackage{graphicx,bm,lineno,amssymb,gensymb}

\graphicspath{{./figures/}}

\usepackage[dvipsnames]{xcolor}
\usepackage[version=3]{mhchem} 
\usepackage{makecell}
\usepackage{array}
\usepackage{float}
\usepackage{graphicx}
\usepackage{subcaption}
\usepackage{adjustbox}
\usepackage{xcolor}
\usepackage{colortbl}

\mciteErrorOnUnknownfalse
\definecolor{mycolor}{RGB}{169, 169, 169}

\title{Towards High-Voltage Cathodes for Zinc-Ion Batteries: Discovery Pipeline and Material Design Rules}

\author{Roberta Pascazio} 
\affiliation{Department of Material Science and Engineering, University of California, Berkeley, CA 94720, U.S.A.}
\alsoaffiliation{Materials Science Division, Lawrence Berkeley National Laboratory, Berkeley, 94720, United States}
\altaffiliation{R.P. and Q.C. contributed equally to this work.}
\author{Qian Chen}
\affiliation{Department of Material Science and Engineering, University of California, Berkeley, CA 94720, U.S.A.}
\alsoaffiliation{Materials Science Division, Lawrence Berkeley National Laboratory, Berkeley, 94720, United States}
\altaffiliation{Present address: Toyota Research Institute of North America, Ann Arbor, Michigan 48105, United States}
\altaffiliation{R.P. and Q.C. contributed equally to this work.}
\author{Haoming Howard Li}
\affiliation{Department of Material Science and Engineering, University of California, Berkeley, CA 94720, U.S.A.}
\author{Aaron D. Kaplan}
\affiliation{Materials Science Division, Lawrence Berkeley National Laboratory, Berkeley, 94720, United States}
\author{Kristin A. Persson}
\affiliation{Department of Material Science and Engineering, University of California, Berkeley, CA 94720, U.S.A.}
\alsoaffiliation{Materials Science Division, Lawrence Berkeley National Laboratory, Berkeley, 94720, United States}
\email{kristinpersson@berkeley.edu}

\date{\today}

\SectionNumbersOn

\begin{document}

\begin{tocentry}
    \centering 
    \includegraphics[width=2in]{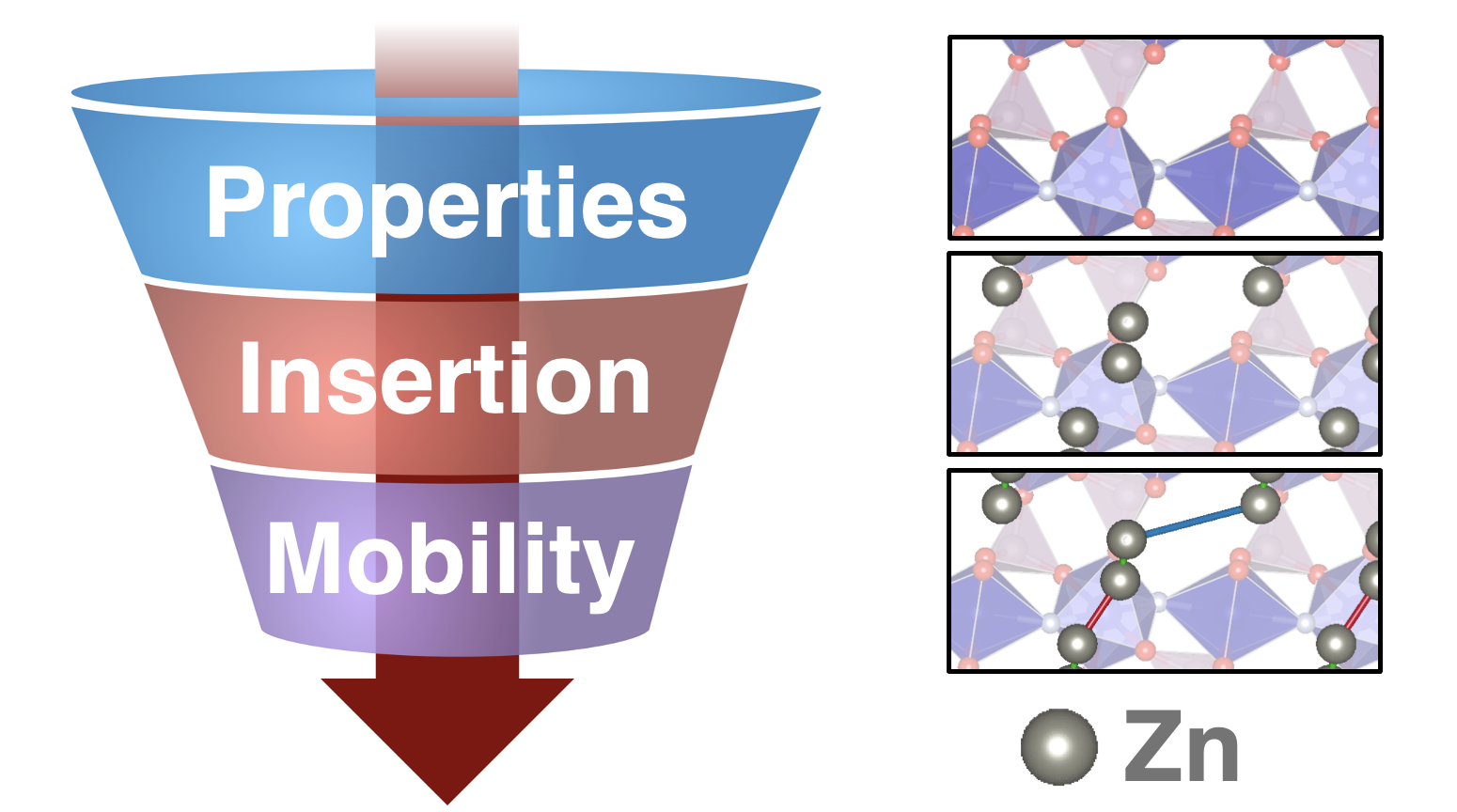}
\end{tocentry}

\begin{abstract}

Efficient energy storage systems are crucial to address the intermittency of renewable energy sources. As multivalent batteries, Zn-ion batteries (ZIBs), while inherently low voltage, offer a promising low cost alternative to Li-ion batteries due to viable use of zinc as the anode. However, to maximize the potential impact of ZIBs, rechargable cathodes with improved Zn diffusion are needed. 
To better understand the chemical and structural factors influencing Zn-ion mobility within battery electrode materials, we employ a high-throughput computational screening approach to systematically evaluate candidate intercalation hosts for ZIB cathodes, expanding the chemical search space on empty intercalation hosts that do not contain Zn.  
We leverage a high-throughput screening funnel to identify promising cathodes in ZIBs, integrating screening criteria with DFT-based calculations of Zn$^{2+}$ intercalation and diffusion inside the host materials. Using this data, we identify the design principles that favor Zn-ion mobility in candidate cathode materials. 
Building on previous work on divalent ion cathodes, this study broadens the chemical space for next-generation multivalent energy storage systems.
\end{abstract}

\section{Introduction \label{sec:intro}}

The interplay between economic development and environmental impact underscores the urgent need for novel sustainable energy sources and, when intermittent renewables fall short of continuous demand, efficient energy storage systems.\cite{WANG2020100078, armand2008, ZHOU2022828}

Since their first commercialization by Sony in the 1990s,\cite{SONY} lithium-ion batteries (LIBs) have dominated the energy storage market, gradually becoming ubiquitous in portable devices and electric vehicles.\cite{lu2011, larcher2015} However, the long-term viability of lithium-ion batteries (LIBs) is hindered by several challenges, including safety risks due to the flammability of commonly used electrolytes \cite{WANG2020100078}, high costs, and the limited availability of critical metals such as cobalt and nickel.\cite{goodenough2010, goodenough2013}
In response to these challenges, multivalent metal-ion batteries (e.g., \ce{Ca^{2+}}, \ce{Mg^{2+}}, \ce{Al^{3+}}) have been suggested as promising alternative energy storage technologies, leveraging the practical use of metal anodes with liquid electrolytes to achieve low cost and competitive volumetric energy density.\cite{rutt2022, MING201958}
As an example, zinc-ion batteries (ZIB) have recently gained attention for the attractive properties of the Zn metal anode which offers (i) a high volumetric energy density (5850 mA h/cm$^3$\cite{MING201958, kundu2018}, compared to $\sim$2000 mA h/cm$^3$ for LIBs\cite{PONROUCH2019253});
and (ii) possible utilization in aqueous batteries\cite{nie2023, song2021} with notable improvement in their safety, sustainability, and operating costs\cite{ju2024, ahn2023}.
When coupled with the appropriate choice of working conditions (e.g., pH) and electrolytes, aqueous ZIBs can be optimized to reduce toxicity and increase the reversibility of the plating/stripping of the Zn anode in candidate ZIBs. To increase their voltage window above the limit imposed by water splitting ($\sim$ 1.23 V):  hybrid aqueous-nonaqueous solvents \cite{yang2020, chang2020, hao2021, li2023} and building on water-in-salt electrolytes, \cite{shen2021}, alternative water-in-organic strategies\cite{tian2023,wang2024} have shown an improvement, effectively suppressing water decomposition and proton intercalation and enabling operational windows of up to $\sim$ 1.6 V vs. Zn/Zn$^{2+}$. Polymer-based\cite{liu2013, sownthari2013} and ionic liquid\cite{abbott2009}-based electrolytes have also been investigated for ZIBs, identifying the formation of Zn dendrites as the major limitation to their cycle lives and electrochemical stabilities.
Nonaqeuous mixtures of Zn salts and organic solvents have also been examined for their electrochemical and transport properties, as well as their charge-transfer performance at the electrode-electrolyte interface. These electrolyte mixtures, particularly those containing acetonitrile and propylene carbonate, displayed reversible deposition on Zn anodes and wide electrochemical windows (up to 3.8 V vs. Zn/Zn$^{2+}$), suggesting their potential application in ZIBs as electrolytes with a variety of cathode materials. \cite{han2016, senguttuvan2016, qiu2019, naveed2019}

Alongside the advancement on electrolytes, the development of high-performance ZIBs has also been directed toward the positive electrode.\cite{ZHOU2022828} 
For instance, a measured capacity of 240 mAh/g at $\sim$1.3 V vs. Zn/Zn$^{2+}$ has been reported for MnO$_2$,\cite{MING201958} leading to high energy densities. 
Promising operated voltages ($\sim$ 1.7 V vs. Zn/Zn$^{2+}$) have also been reported for a wide variety of materials, such as Prussian blue analogues (PBAs) and organic electrodes, albeit with the trade-off of reduced stability over repeated cycles and/or lower volumetric capacities.\cite{kundu2018} \\

However, multivalent ions, and Zn$^{2+}$ among them, are known to strongly interact with water and with the host electrode materials, leading to slower diffusion which, in turn, correlates with  reduced reversibility and diminished cyclability. 
In all divalent intercalating electrode materials, mobility of the active ion is known to a primary concern.\cite{rutt2022, kim2024_1} Hence,  designing improved ZIB cathodes will require a deep understanding of Zn mobility. Overall, Zn mobility in ZIBs is known to be primarily influenced by two key characteristics\cite{jun2024}: (i) chemical factors, including electronegativity, which directly affects the covalent or ionic nature of bonds between Zn and the surrounding ionic framework\cite{manthiram1989, etourneau1992}; and (ii) structural and topological factors, such as the local bonding environment where stronger bonds may introduce energetic barriers to diffusivity\cite{liu2015} and the dimensionality of percolating channels for Zn diffusion.\cite{MING201958}.
Previous research on Zn-ion battery materials have investigated the influence of these factors, employing them as design rules for the morphological and structural engineering of the diffusion pathways.\cite{kundu2018} Zn mobility has also been investigated computationally through screenings of its preferred coordination in both activated and stable sites within different candidate host structures\cite{canepa2017}, as well as by systematic studies of Zn intercalation sites and pathways within host materials of diverse chemical compositions\cite{liu2015, miranda2024, gautam2015, canepa2017, liu2016, rong2015}. However, these studies have typically focused on incorporating or replacing the working ions into known crystallographic sites within host structures. To date, there have been no extensive screening efforts involving candidate cathode materials that do not contain Zn or other working ions in their as-synthesized state. In contrast, experimental examples of such 'empty-host' Zn cathodes are birnessite-MnO$_2$\cite{GUO2020100396}, for which traces of elemental Zn are confirmed by energy-dispersive X-ray (EDX), and V$_2$O$_5$ polymorphs, which have been investigated in the pre-inserted M$_x$V$_2$O$_5$ form (where M = alkali metal, and  1.1 $\leq$ x $\leq$ 1.2). \cite{zhang2021}
Hence, screening efforts should include empty intercalation hosts, which often outperform structures containing the migrating ion,\cite{rutt2022, rong2015} as they often present the active ion in a deep electrostatic potential well which in turn results in poorer mobility.

In this work, we build upon previous works on Mg\cite{rutt2022} and Ca\cite{kim2024_1} cathodes, employing existing automated computational infrastructure\cite{rutt2022, li2024} to expand our understanding of the chemical space for divalent ion batteries and to identify candidate high-voltage multivalent cathodes for ZIBs. We first motivate screening criteria to select the most promising intercalation hosts, followed by high-throughput DFT calculations to explore the intercalation of Zn$^{2+}$. We maintain the possibility for aqueous applications by integrating a well-established screening criterion to evaluate candidate stability against passivation and corrosion in aqueous environments, while also predicting the composition of the resulting passivation products.\cite{singh2017, ong2008} We then investigate ion mobility in the most promising Zn cathodes, highlighting the factors that influence Zn diffusivity.

\section{Methods: Cathode discovery pipeline \label{sec:methods}}
\begin{figure}[b!]
     \centering
     \includegraphics[width=\textwidth]{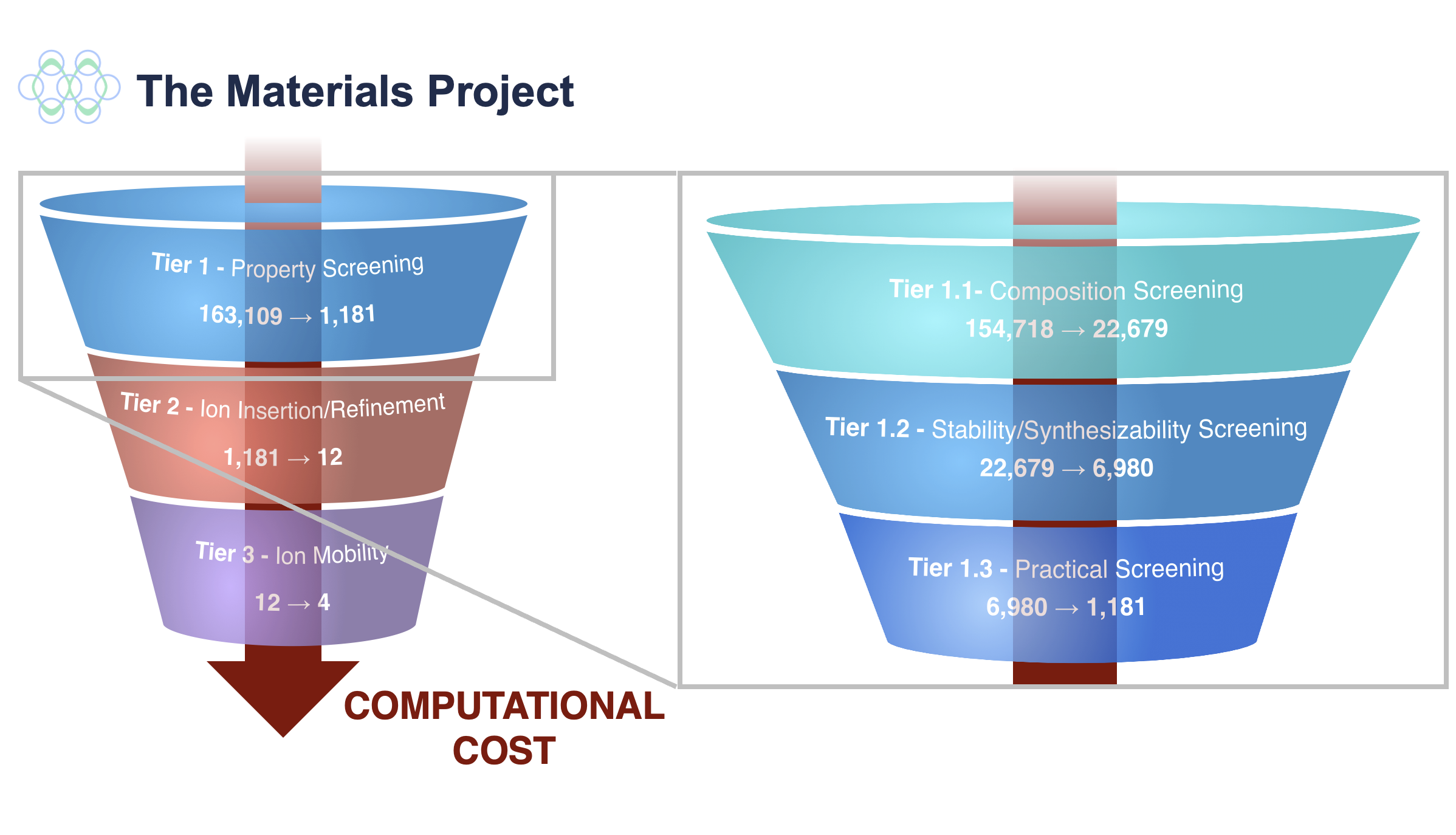}    
     \caption{(left) Funnel diagram summarizing the screening process for identifying cathodes for ZIBs from host candidates in the deintercalated state, showing the number of materials entering and exiting each stage. \\ 
     The process is divided into three stages, ordered by increasing computational cost: \\
     \textbf{Tier 1.} Materials Project property screening; \\ \textbf{Tier 2.} Ion insertion calculations and additional property screenings/prototype matches; \\ \textbf{Tier 3.} Ion mobility calculations. \\ (right) Funnel diagram summarizing the screening sub-tiers for \textbf{Tier 1} in the screening funnel. The process is divided into three stages: \\ \textbf{Tier 1.1.} Composition screening; \\ \textbf{Tier 1.2.} Stability/Synthesizability screening; \\ \textbf{Tier 1.3.} Practical Screening.}
     \label{fig:funnel}
\end{figure}

The screening funnel, represented on the left in Figure \ref{fig:funnel}, follows a similar strategy to previous searches\cite{li2024, rutt2022} for multivalent cathodes, in which successive tiers of the funnel require increasingly demanding calculations.
This approach helps minimize the use of more resource-intensive methods, applying them only to a smaller set of promising candidates.

\subsection{Tier 1: Property Screening \label{sec:tier1}}
In the initial  Tier 1 screening, we select materials with desirable properties from the 2023.11.1 version of the Materials Project (MP) database\cite{jain2013}, which contains 163,109 materials.
The property screening is then divided into 3 sub-tiers, represented on the right in Figure \ref{fig:funnel}. 

Further details on the screening tiers, including specifics and metrics on the applied filters, are provided in Sec. \ref{sec:tier1_SI} of the Supporting Information (SI) of this manuscript.
In this Tier, candidates were assessed as follows. \textbf{(1.1) Composition Screening}: Exclusion of chemically a priori undesirable compositions, which includes precious metals, radioactive, toxic and redox-inactive elements. To focus our effort on empty host materials, in this tier, we also exclude Zn or other known working ions to simplify the evaluation of ion mobility and diffusivity. This screening narrowed the pool of potential candidates from 163,109 to 22,769 structures; \textbf{(1.2) Stability/Synthesizability Screening}: Exclusion of materials that may be too unstable to be synthesized and/or survive in the specific electrochemical working conditions for high-voltage ZIBs. This filter includes the evaluation of known descriptors for thermodynamic stability, such as the energy above the hull (the distance of a phase from the convex energy hull of its most stable phases)\cite{hautier2012,sun2016} as well as a screening for reasonable thermodynamical stability against dissolution, passivation and/or corrosion in the event of aqueous applications.
In total, this filter reduced the number of candidates from 22,769 to 6,980 viable materials.\textbf{(1.3) Practical Screening}: Selection of cost-effective and high-performance materials. This screening tier excluded elements with large cost to capacity ratio, as well as structural frameworks presenting similar crystal structure types and different transition metal (TM) ratios (e.g., NASICON structures (TM)$_2$(PO$_4$)$_3$ with different TM combinations). This final screening tier reduced the optimal candidates from 6,980 to 4,297 structures, belonging to 1,181 distinct crystal structure types.
Overall, this process narrowed the initial 163,109 structures to 1,181 candidates.

\subsection{Tier 2: Ion insertion \label{sec:tier2}}
To identify potential intercalation sites for Zn$^{2+}$, an insertion algorithm\cite{shen2020} based on DFT-calculated electronic charge densities was employed to insert Zn$^{2+}$ ions in empty host structures. The algorithm was repeated for multiple Zn insertions until (i) the transition metal element in the resulting intercalated structure reaches the lowest available oxidation state; or (ii) a structural mismatch between the intercalated and empty host structures occurs, indicating non-topotactic intercalation\cite{li2024}; or (iii) the new structure is rendered too unstable; or (iv) the volume change is larger than 20\%. This tier required the calculation of several DFT-calculated properties for the intercalated structures, including stability, average intercalation voltage, energy density, and the optimized inserted structure, making it computationally expensive and hence impractical for large material databases. For this reason, we restricted the ion insertion calculations to candidate structures containing common high-voltage TMs (Mn, Co, Cr and Ni). Out of the 1,181 candidates, the ion insertion calculations were successfully completed for 313 of them.

\subsubsection{Additional Screening and Prototype Matching \label{sec:structure_match}}
The top-performing intercalated materials from the insertion electrode calculations were further screened based the properties calculated in Tier 2, selecting high-performance materials (presenting high average intercalation voltages, gravimetric capacities and energy densities $>$ 300 Wh/kg). The materials, as a function of state of charge, were then filtered for stability against conversion reactions, \cite{hannah2018} using data from the Materials Project phase diagram\cite{gunter2012}, \texttt{pymatgen}\cite{ONG2013314}.  A more detailed description of the filtering criteria is provided in the SI.
These conditions, not present in previous cathode pipelines \cite{rutt2022,kim2024_1}, ensured that candidate materials were thermodynamically stable, and exhibited reasonable protection against dissolution in aqueous media \cite{MING201958, singh2017}.
This screening reduced the 313 candidates obtained in Tier 1 to the 37 best performing materials. 
For the last tier of calculations, priority was given to candidates whose structural frameworks closely match known synthesized materials among the 37 best performers.
The framework assessment was conducted by matching structures to those in the 5.3.0 version of the Inorganic Crystal Structure Database (ICSD) via \texttt{pymatgen}
through both exact and ``looser'' structure matches, accounting for structural disorder and doping.
In the subset structure match, we permitted matches between candidate and ICSD materials upon substitution of candidate TM sites with other isovalent TMs, and/or substitution of chalcogen and halide sites with other members of their respective groups.
The structure matching between the structures was performed using the default tolerance factors.\cite{li2024} In this final screening tier, out of the 37 top candidates, 12 met the ``subset structure match'' criteria, indicating close structural similarity to experimentally synthesized materials. These 12 candidates were then subjected to mobility calculations, which are detailed in the following section.

\subsection{Tier 3: Ion mobility \label{sec:tier3}}
In the third tier of the screening funnel, the working ion sites of the top 12 performers were used to construct a \text{MigrationGraph}\cite{shen2023}, mapping the interconnected network of metastable ion sites through a series of ``hops" \cite{li2024}. This step identifies symmetrically equivalent sites and hops, generates potential migration pathways, and collects them into a periodic \texttt{MigrationGraph} document\cite{koistinen2017}.
Compounds without periodically repeatable Zn migration were discarded, limiting this computationally expensive step to only the candidates displaying feasible migration.

As an approximation of Zn$^{2+}$ mobility, we used ApproxNEB\cite{rong2016} to evaluate the energy profile of the migration pathways.
ApproxNEB, as implemented in the \texttt{atomate}\cite{kiran2017} and \texttt{atomate2}\cite{ganose2025} packages, offers a robust and efficient alternative to NEB \cite{mills1995neb} by decoupling images along the reaction coordinate and replacing the NEB spring forces with constrained relaxation of the ionic coordinates.
In this work, all ApproxNEB calculations were conducted using a structure within the deintercalated/dilute limit, with only one Zn ion in the simulation supercell. To accurately represent the dilute limit and avoid self-interaction effects between Zn ions in neighboring periodic cells, the intercalated structures were generated using \texttt{pymatgen}\cite{ONG2013314} to ensure that periodic Zn images are at least 7 \AA{} apart. The structures were then relaxed using DFT with the working ion and its antipodal site fixed, yielding migration energies. 
In particular, the energy barriers associated with the hops between sites were calculated and mapped onto their respective migration graphs, calculating the shortest percolating pathways through Dijkstra’s algorithm. 
ApproxNEB is known to overestimate migration energy (E$_m$) values compared to NEB, as it is not guaranteed to yield images on the minimum energy path of the potential energy surface, unlike NEB \cite{rong2016, rutt2022, li2024, kim2024_1}.
For this reason, ApproxNEB energy barriers provide an upper limit to the barrier in our screening process: such that, assuming nanosized materials, a migration energy threshold of $\sim 1$ eV is applied.\cite{rong2015}.

In this screening protocol, 9 of the ApproxNEB workflows obtained for the 12 top candidates identified via the ICSD subset match successfully converged, revealing possible pathways for Zn$^{2+}$ migration.
In particular, 4 materials exhibited a percolation barrier below the 1 eV threshold for at least one migration pathway, and have thus been identified as promising in terms of ion mobility and synthetic viability. 
The details and energetic landscapes of the calculated pathways will be the subject of the following section. 
Once promising materials are identified through ApproxNEB calculations, further in-depth diffusion analyses -- such as climbing image (CI)-NEB calculations\cite{henkelman2000} and \textit{ab initio} molecular dynamics (AIMD) -- can be employed to gain a more comprehensive understanding of the energetics and morphology of the migration network. 

\section{Results \& Discussion \label{sec:results}}
\subsection{Screening Results\label{sec:screen_results}}
Here, we expand on the results obtained from the screening pipeline, highlighting the key findings obtained from the analysis of the most promising candidates.

Of the candidates identified in Tier 1, 88.2\% are filtered out by the Tier 2 criteria: 21.7\% due to low average theoretical intercalation voltages, 62.6\% due to stability requirements, and 42.5\% with below-threshold gravimetric capacities or energy densities.

\begin{figure}[t!]
    \centering
        \includegraphics[width=\linewidth]{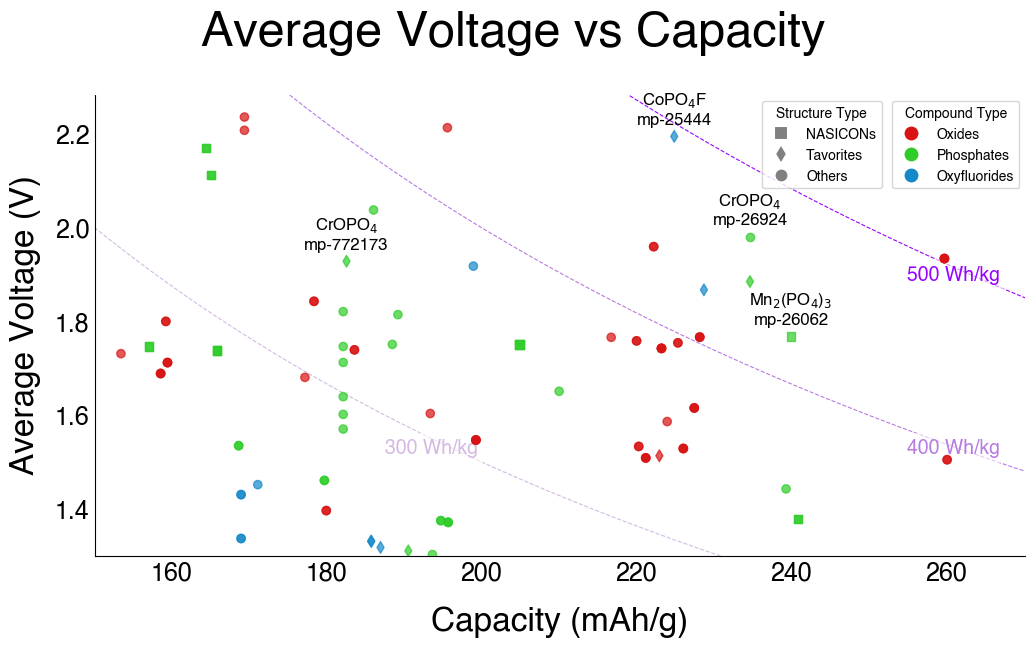}
        \caption{Distribution of best-performing candidate materials resulting from Tier 2 by average voltage and theoretical capacity.}
        \label{fig:res}
\end{figure}

\begin{figure}[t!]
        \centering
        \includegraphics[width=\linewidth]{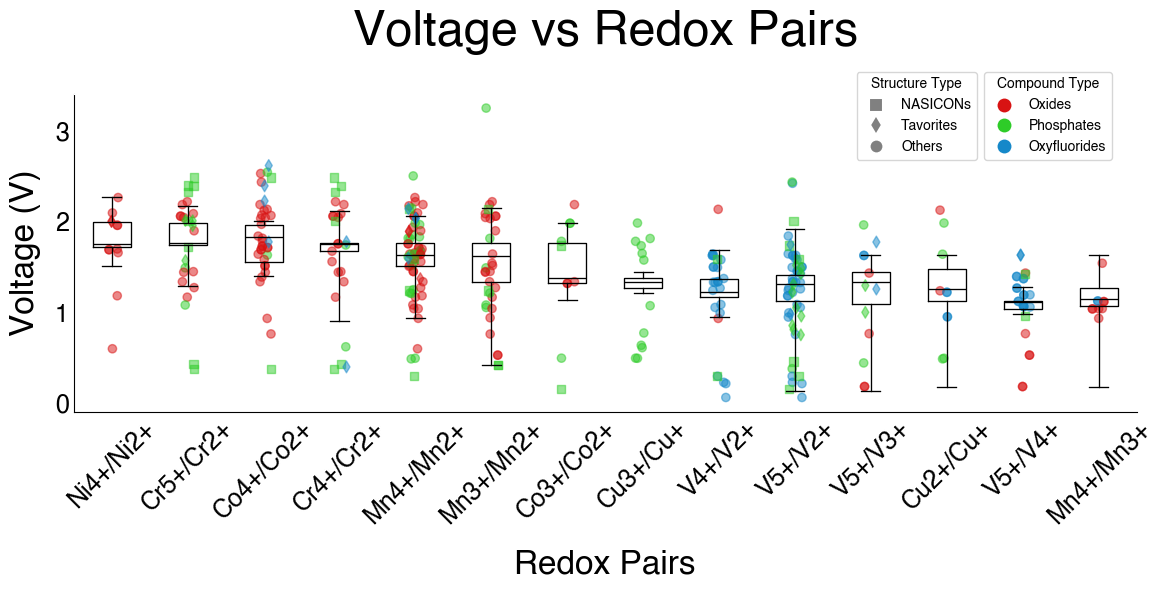}
        \caption{Voltage distribution (max and min voltages) of the best candidates as a function of the most common redox pairs in the host structures. The statistical analysis was conducted on the average voltages of the candidates, which are plotted in decreasing order of median average voltage.}
    \label{fig:res2}
\end{figure}

Figures \ref{fig:res} and \ref{fig:res2}, summarize the outputs of Tier 2, representing the overall performance (Fig. \ref{fig:res}) and voltage windows (Fig. \ref{fig:res2}) of the top candidates (gravimetric energy density $>$ 300 Wh/kg) resulting from the ion insertion step, including the 4 best candidates obtained at the end of the screening procedure (Figure \ref{fig:res}). However, we find that the average intercalation potentials of the top candidates exceed the oxygen evolution reaction (OER) potential, which occurs at approximately 0.9–0.94 V vs. SHE ($\sim$0.1 V vs. Zn/Zn$^{2+}$) in the pH range relevant to aqueous applications (pH~=~5–5.5). This suggests that these materials may only be amenable for aqueous applications with specialized electrolytes, e.g. water-in-salt WISE electrolytes. ~ \cite{suo2015, borodin2020}

Based on the screening results of Fig. \ref{fig:res2}, and in accordance with known chemical trends,\cite{hautier2011, li2017}  oxyfluorides and phosphates exhibit higher voltages compared to oxides. In particular, the candidates with the highest voltages were predominantly polyanion compounds, specifically (fluoro)phosphates. This trend conforms with the well-established strong inductive effect of polyanion groups and is commonly observed in both Li-ion \cite{hautier2011, li2017} and Na-ion \cite{ni2017} batteries, where (fluoro)phosphate cathode materials typically exhibit very high voltages.
Similar considerations can be made in terms of the redox center. Figure ~\ref{fig:res2} demonstrates that cations such as Ni$^{4+}$, Co$^{4+}$ and Cr$^{5+}$ exhibit higher voltage distributions, which correlates with heavier redox-active cations within the same period or those with higher oxidation states.~\cite{liu2016_1, li2024_1}
A more detailed description of the effect of the polyanions and the redox centers on the potential of the host materials is provided in Section \textbf{Structural and Chemistry Effects}. \\
The results demonstrate the effectiveness of our screening protocol, as the targeted approach significantly narrows the candidate pool, restricting computationally expensive ion diffusion calculations to only the most promising host materials. 
Thirty seven candidates — around 12\% of the 313 electrode candidates which underwent insertion calculations — met the additional screening criteria in the second phase of Tier 2. This selection was further refined by the prototype matching, yielding 12 final candidates (about 4\% of the initial pool).

\begin{figure}[t]
    \centering
    \renewcommand{\arraystretch}{1.2}
    \setlength{\tabcolsep}{7pt}
    \setlength{\arrayrulewidth}{1pt}

    \begin{tabular}{l!{\color{mycolor}\vrule width 1pt} l}
        \begin{subfigure}[t]{0.49\textwidth}
            \centering
            \includegraphics[width=\linewidth]{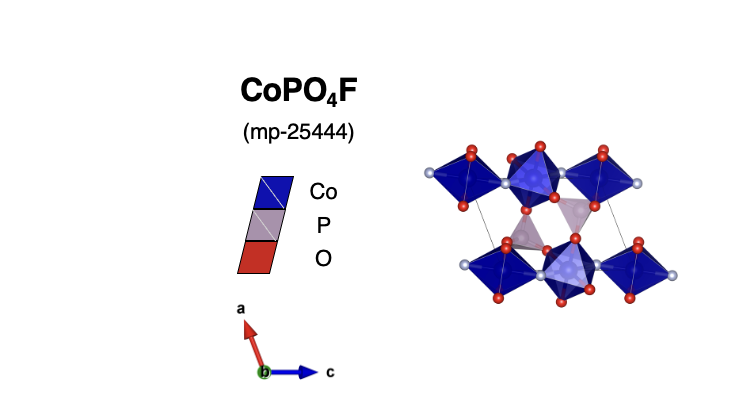}
            \caption{}
            \label{fig:25444_candidate}
            \vspace{1em} 
        \end{subfigure} 
        &
        \begin{subfigure}[t]{0.49\textwidth}
            \centering
            \includegraphics[width=\linewidth]{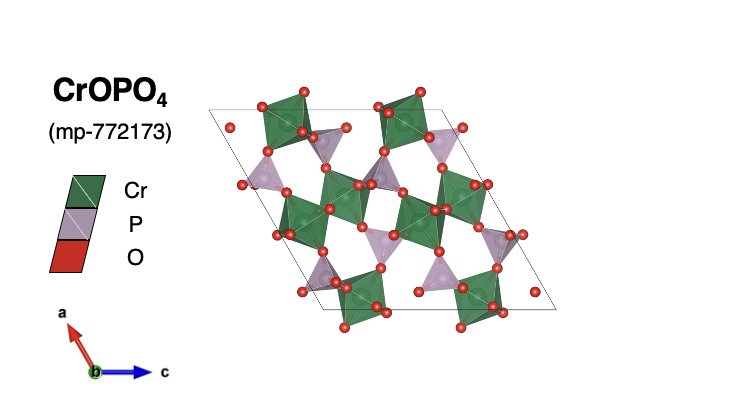}
            \caption{}
            \label{fig:772173_candidate}
            \vspace{1em} 
        \end{subfigure} 
        \\
     \arrayrulecolor{mycolor}\hline
        \begin{subfigure}[t]{0.49\textwidth}
            \centering
            \includegraphics[width=\linewidth]{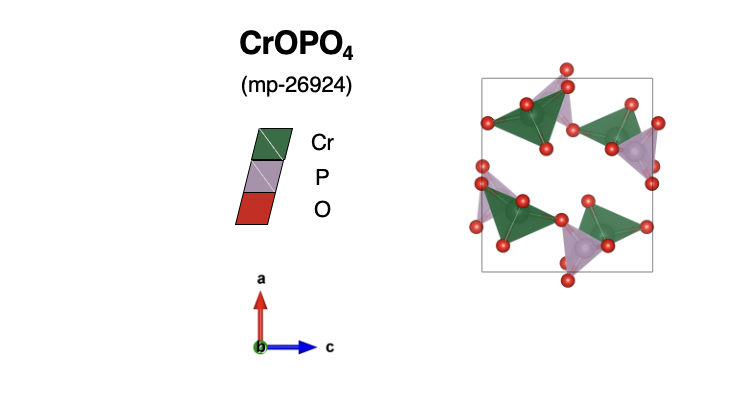}
            \caption{}
            \label{fig:26924_candidate}
        \end{subfigure} 
        &
        \begin{subfigure}[t]{0.49\textwidth}
            \centering
            \includegraphics[width=\linewidth]{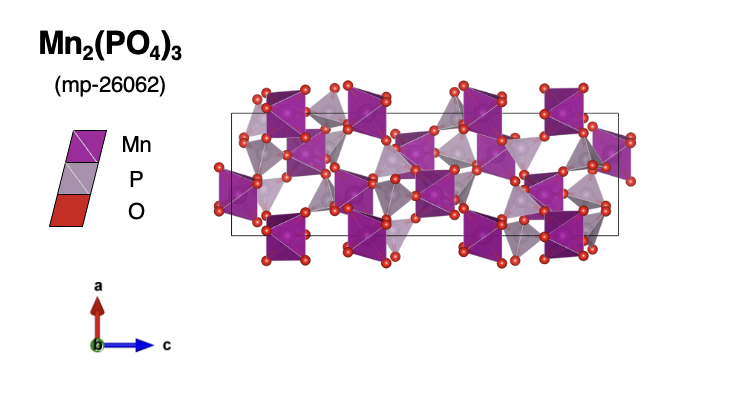}
            \caption{}
            \label{fig:26062_candidate}
        \end{subfigure} 
    \end{tabular}

    \caption{Unit cell crystal structures of the best performing candidates obtained from the screening pipeline: (\ref{fig:25444_candidate}) triclinic tavorite CoPO$_4$F (mp-25444), (\ref{fig:772173_candidate}) monoclinic tavorite CrOPO$_4$ (mp-772173), (\ref{fig:26924_candidate}) orthorhombic Cr phosphate CrOPO$_4$ (mp-26294), (\ref{fig:26062_candidate}) trigonal NASICON Mn$_2$(PO$_4$)$_3$ (mp-26062).}
    \label{fig:candidates}
\end{figure}

As is discussed in Section \textbf{Structural and Chemistry Effects}, various polymorphs (e.g. CoPO$_4$F\cite{fedotov2016, mueller2011}, MnP$_2$O$_7$\cite{li2024}) and/or compositions of some of these 12 candidates have been investigated as intercalation electrodes in previous work on known cathode prototypes for Zn and other working ions (e.g., the study of Na intercalation/deintercalation mechanism in doped equivalents of Mn$_2$(PO$_4$)$_3$ such as MnV(PO$_4$)$_3$ \cite{nisar2018} and ofMn$_3$V$_2$(PO$_4$)$_3$ in ZIBs: \cite{li2016}
However, the specific compositions of these materials have to our knowledge not been considered for Zn ion intercalation.  Furthermore, our approach also revealed several promising candidates that have not yet been explored in literature (e.g. Mn$_2$(PO$_4$)$_3$), highlighting the potential of the employed pipeline for the discovery of promising novel materials.
Following the final ApproxNEB step in the pipeline, four out of twelve candidates were selected, as they demonstrated the most favorable migration pathways and energetics. The four best performing candidates resulting from the overall cathode discovery pipeline are represented in Figure \ref{fig:candidates}.
Their IDs, calculated properties and ICSD prototypes are represented in Table \ref{tab:tab1}.

\begin{table}[h!]
  \centering
  \renewcommand{\arraystretch}{2.2}
  \resizebox{\textwidth}{!}{%
    \begin{tabular}{c c c c c c c c c c} \hline
      \makecell{Formula \\ mp-id} & Symmetry & \makecell{Pourbaix \\ $\Delta$G$_{\text{max}}$ \\ (meV)} & \makecell{Intercalation \\ Voltage (V)} & \makecell{$\Delta$V \\ (\%)} & \makecell{Gravimetric \\ Capacity \\ (mAh$/$g)} & \makecell{Gravimetric \\ Energy Density \\ (Wh$/$kg)} & \makecell{Charge \\ Stability \\ (meV$/$atom)} & \makecell{ApproxNEB Barrier \\ (meV)} & Prototype \\ \hline      \makecell{CoPO$_4$F \\ (mp-25444)} & \makecell{P$\overline{1}$ \\ (triclinic)} & 0.30 & 2.19 & 6\% & 224.93 & 493.60 & 0.10 & 772 (1D) (2D with 1.2) & \makecell{VOPO$_4$ \\ (mp-1216303)},
    \makecell{SbOPO$_4$ \\ (mp-9750)} \\
    \makecell{CrOPO$_4$ \\ (mp-772173)} & \makecell{P2$_1/$c \\ (monoclinic)} & 0.34 & 1.93 & 6\% & 182.57 & 352.01 & 0.10 & 958 (1D) & \makecell{NbOPO$_4$ \\ (mp-542453)}  \\
      \makecell{CrOPO$_4$ \\ (mp-26924)} & \makecell{Pnma \\ (orthorhombic)} & 0.30 & 1.98 & 4\% & 234.71 & 464.82 & 0.05 & 774 + 951 (1D) & \makecell{VOPO$_4$ \\ (mp-25265)} \\
      \makecell{Mn$_2$(PO$_4$)$_3$ \\ (mp-26062)} & \makecell{R$\overline{3}$ \\ (trigonal)} & 0.45 & 1.77 & 20\% & 240.02 & 424.23 & 0.06 & 894 (3D) & \makecell{Nb$_2$(PO$_4$)$_3$ \\ (mp-17242)} \\
  \hline
    \end{tabular}%
  }
  \caption{Summary of Electrode Properties for the Four Best Performing Zn Cathodes.}
  \label{tab:tab1}
\end{table}
We proceed to analyze the results of the calculations performed in Tier 2 and Tier 3, identifying the topology of the ion insertion sites and of the corresponding percolation pathways. 

\subsection{ApproxNEB results \label{sec:approxneb_results}}
As discussed in Section \textbf{Tier 3: Ion mobility}, efficient kinetics are crucial for utilizing Zn as a working ion in prospective battery materials. Therefore, we employed ApproxNEB calculations to evaluate Zn migration pathways for the top candidates identified in Tier 3.
We define a percolating pathway as the trajectory from a Zn site within a unit cell to a periodic site in an adjacent unit cell. Each pathway comprises symmetrically distinct hops between intercalation sites, or endpoints. The endpoints are labeled as A, B, etc., in order of increasing destabilization (i.e. insertion energy), with numerical subscripts distinguishing different endpoints of the same energy. Periodic images are indicated using a superscript prime symbol $^\prime$. We emphasize that the pathways are analyzed in the dilute lattice limit, as these materials are synthesized in their charged state. While low barriers in this limit are a necessary criterion, they are not sufficient, as cation-cation interactions can hinder mobility and reduce rate capability in the partially discharged state. 

\begin{figure}[t!]
    \centering
    \renewcommand{\arraystretch}{1.2} 
    \setlength{\tabcolsep}{7pt}
    \setlength{\arrayrulewidth}{1pt}

    \begin{tabular}{l!{\color{mycolor}\vrule width 1pt} l}
          \begin{subfigure}[t]{0.49\textwidth}
         \centering
         \includegraphics[width=\linewidth]{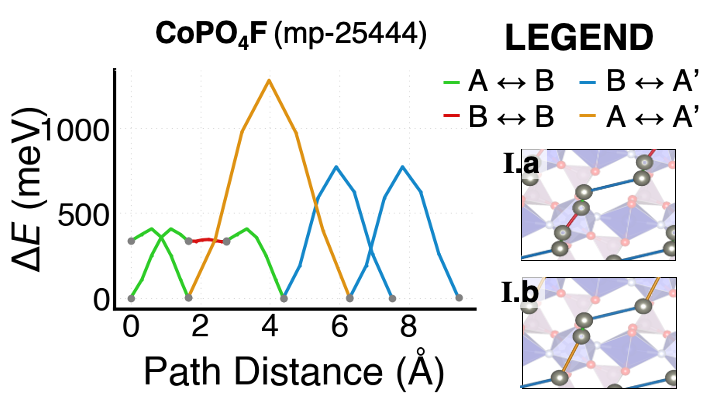}
         \caption{}
         \label{fig:25444_pathways}
         \vspace{1em} 
     \end{subfigure}
 & 
     \begin{subfigure}[t]{0.49\textwidth}
         \centering
         \includegraphics[width=\linewidth]{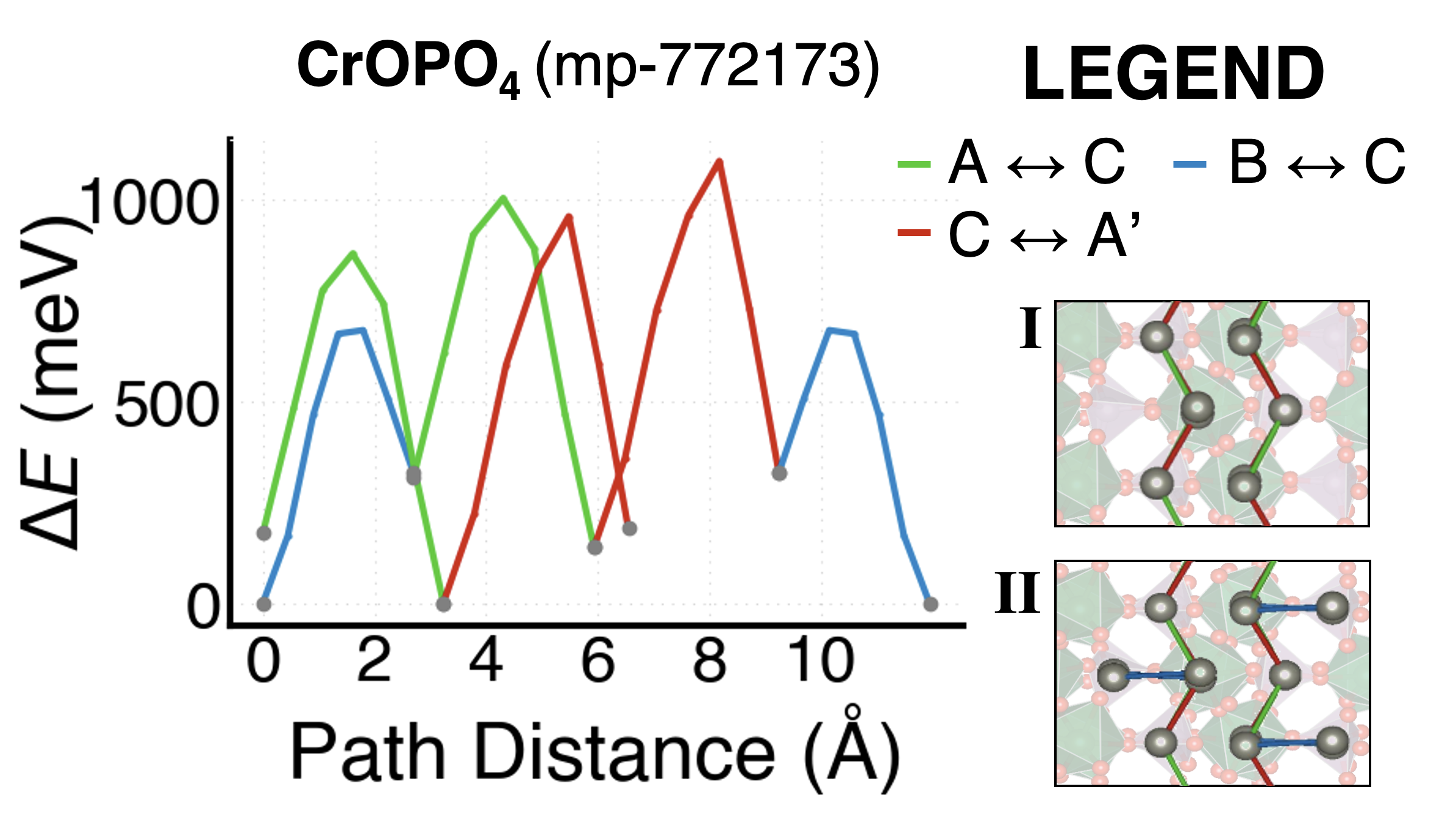}
         \caption{}
         \label{fig:772173_pathways}
         \vspace{1em}
     \end{subfigure}

        \\
     \arrayrulecolor{mycolor}\hline

     \begin{subfigure}[t]{0.49\textwidth}
         \centering
         \includegraphics[width=\linewidth]{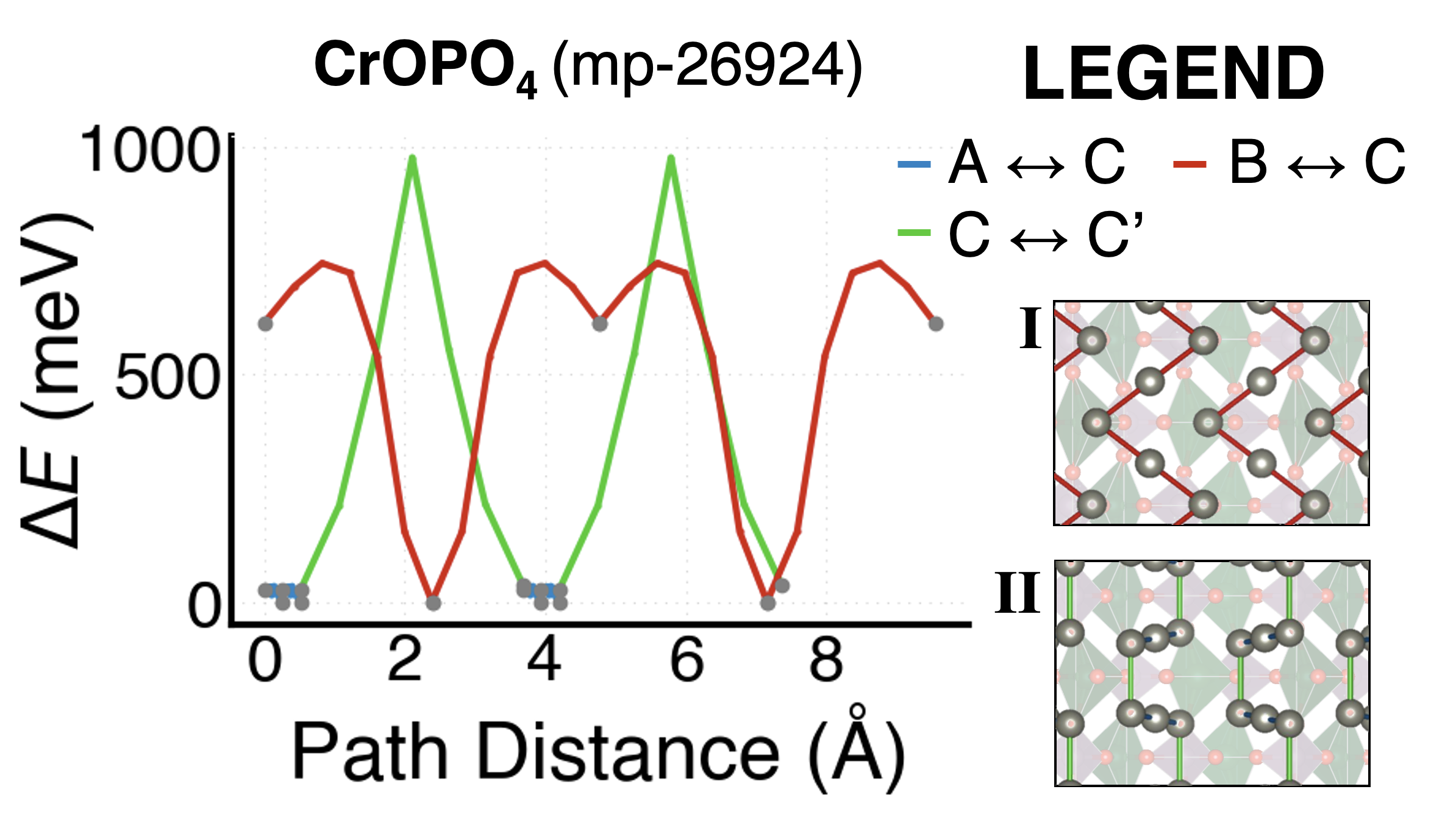}
         \caption{}
         \label{fig:26924_pathways}
     \end{subfigure}
&
     \begin{subfigure}[t]{0.49\textwidth}
         \centering
         \includegraphics[width=\linewidth]{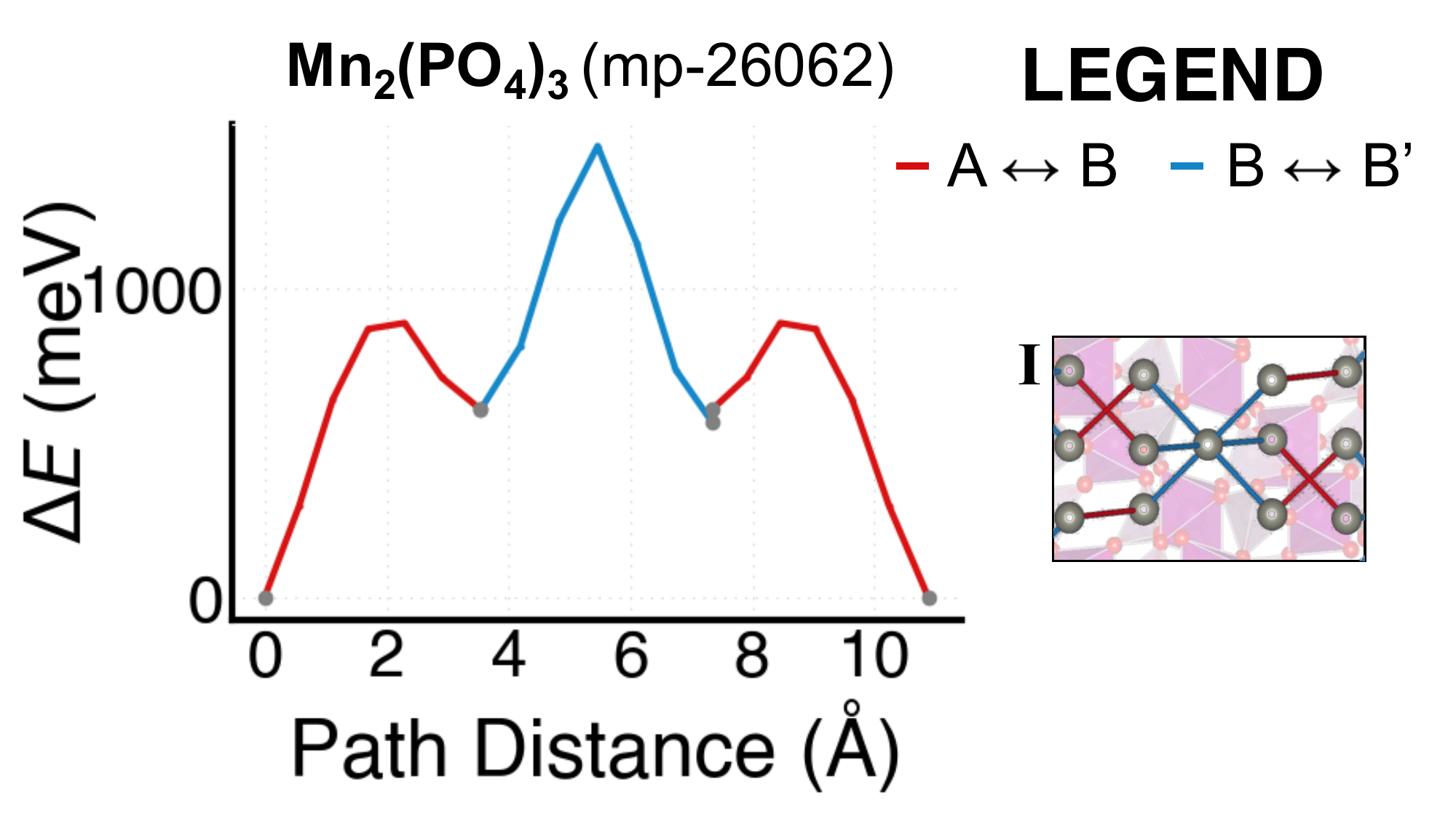}
         \caption{}
         \label{fig:26062_pathways}
     \end{subfigure}   
    \end{tabular}
     \caption{Energy landscape plots for Zn$^{2+}$ migration along percolating pathway with the lowest kinetic barriers in the four candidate materials. Each hop in the energy profile (left) is mapped with the same color in the pathway (right) and labeled through its endpoints, which are defined as symmetrically unique positions through the notation in this section. \\(\ref{fig:25444_pathways}) Triclinic tavorite CoPO$_4$F (mp-25444), exhibiting energetic barriers of 772 meV and 1283 meV over total path distances of respectively 7.51 \ \AA{} and 9.43 \ \AA{} \ (both \textbf{PathwayI}, left and right). (\ref{fig:772173_pathways}) Monoclinic tavorite CrOPO$_4$ (mp-772173), both with an energetic barrier of 958 meV, across total path distances of respectively 6.55 \AA{} \ (\textbf{PathwayI}, on the left) and 11.95 \AA{} \ (\textbf{PathwayII}, on the right). (\ref{fig:26924_pathways}) Orthorhombic Cr phosphate CrOPO$_4$ (mp-26924), showing energetic barriers of 774 meV and 951 eV along a total path distance of respectively 9.55 \AA{} \ (\textbf{PathwayI}, on the left) and 7.39 \AA{} \ (\textbf{PathwayII}, on the right). (\ref{fig:26062_pathways}) Trigonal NASICON Mn$_2$(PO$_4$)$_3$ (mp-26062), with an energetic barrier of 894 meV over a total path distance of 10.88 \AA.}
     \label{fig:pathways}
\end{figure}

Overall,
four of the candidates presented a barrier below the 1 eV threshold for at least one percolation pathway.
The ApproxNEB results for these materials show the following energy barriers for Zn migration: 772 meV for CoPO$_4$F (mp-25444), 958 meV for CrOPO$_4$ (mp-772173), 951 meV for CrOPO$_4$ (mp-26924), and 894 meV for Mn$_2$(PO$_4$)$_3$ (mp-26062). In Figure \ref{fig:pathways}, we  present the calculated energy profiles for the the most kinetically favorable pathways for long-range Zn migration in the four best candidates. The screening results indicate that CoPO$_4$F (mp-25444) shows potential as a high-energy density ZIB cathode. However, nanosizing and testing under repeated cycling conditions is needed as the limited Zn$^{2+}$ mobility could lead to reduced stability and poor rate capability over time.\cite{canepa2017_1}

\subsubsection{CoPO$_4$F (mp-25444) \label{sec:copo4f}}
As shown in Table \ref{tab:tab1}, CoPO$_4$F (mp-25444) exhibits an average theoretical Zn intercalation voltage of 2.19 V vs Zn/Zn$^{2+}$, and a volume change of 6\%. It should be noted that ZnCoP$_4$F is close to the limit of our stability criteria, with the fully intercalated state (ZnCoP$_4$F) exhibiting an energy above hull of 100 meV/atom and its charged phase at 121 meV/atom.  According to the Materials Project phase diagram,\cite{jain2013, ong2008} ZnCoP$_4$F is predicted to decompose into Zn$_3$(PO$_4$)$_2$, Co$_3$(PO$_4$)$_2$ and CoF$_2$, indicating that full theoretical 224.93 mAh/g  gravimetric capacity and 493.60 Wh/kg energy density may not be achievable.  For aqueous electrolyte applications, CoPO$_4$F  is 300 meV/atom unstable against P(OH)$_2$$^-$, F$^-$ and Co(OH)$_2$, which means that a Co(OH)$_2$ surface passivation layer may form in aqueous applications.\cite{singh2017}
 
CoPO$_4$F crystallizes in the low-symmetry triclinic P$\overline{1}$ space group, with calculated lattice parameters of 
a = 5.27 \AA, b = 5.30 \AA, c = 7.39 \AA{} $ $ and relative angles of $\alpha$ = 108.57\degree, $\beta$ = 107.81\degree, $\gamma$ = 95.60\degree. This candidate belongs to the tavorite family: its structure features two inequivalent Co sites, both forming CoO$_4$F$_2$ octahedra that share corners with two other CoO$_4$F$_2$ octahedra and four equivalent PO$_4$ tetrahedra.

In this candidate, Zn can be inserted into one of two symmetrically unique positions, labeled A and B. The B site corresponds to the least stable Zn configuration, as it is located closer to a cation (the Co atom in a CoO$_4$F$_4$ octahedron). 
ApproxNEB calculations confirm a main one-dimensional (1D) pathway characterized by three symmetrically unique hops (Figure \ref{fig:25444_pathways}): a 407 meV hop between adjacent A and B sites (A$_1$ and B$_1$), a 13 meV hop between two B sites (B$_1$ and B$_2$), another 407 meV hop between adjacent A and B sites (this time in the reverse direction, i.e. between B and an adjacent A site), and a 772 meV hop between two A sites (A$_2$ and A$_1$), terminating in a different unit cell. This pathway, named \textbf{PathwayI.a}, thus follows the sequence:
A$_1 \rightarrow$ B$_1 \rightarrow$ B$_2 \rightarrow$ A$'_2 \rightarrow$ A$'_1$. An alternative sequence of hops for \textbf{PathwayI.a} could involve replacing the two intermediate hops with a 448 meV hop between the A$_2$ site and the non-adjacent B$_2$ site, leading to an overall A$_1 \rightarrow$ B$_2 \rightarrow$ A$'_2 \rightarrow$ A$'_1$ with the same dimensionality that terminates in a periodic A$_1$ position in a separate unit cell. This alternative sequence \textbf{PathwayI.b}), though highly hindered (with a 1283 meV hop between the A$_1$ and A$_2$ sites), could extend the system's network to two dimensions. A 2D pathway would be desirable in order to reduce the probability that defects and impurities block Zn migration.

\subsubsection{CrOPO$_4$ (mp-26924, mp-772173) \label{sec:cropo4}}
In our cathode screening pipeline, two polymorphs of CrOPO$_4$ were identified as potential candidates for Zn-ion cathodes: a monoclinic polymorph, mp-772173, and an orthorhombic polymorph, mp-26924.
Table \ref{tab:tab1} highlights the calculated electrochemical properties of the two polymorphs. The first polymorph, mp-772173, displays a higher theoretical intercalation voltage (1.98 V vs Zn/Zn$^{2+}$), greater energy density (464.82 Wh/kg), and a smaller volume change (4\%) compared to mp-26924, which has an intercalation voltage of 1.93 V vs Zn/Zn$^{2+}$, an energy density of 352.01 Wh/kg, and a 6\% volume change. The monoclinic polymorph belongs to the P2$_1/$c space group with lattice parameters of a = 12.68 \AA, b = 5.04 \AA, c = 12.80 \AA{} $ $ and relative angles of  108.57\degree, $\alpha$ = $\gamma$ = 90.00\degree, $\beta$ = 119.81\degree. The structure of this polymorph is less stable, with a charge state energy above hull of 100 meV/atom against CrO$_2$, CrPO$_4$, ZnCr$_2$O$_4$ and Zn$_3$(PO$_4$)$_2$. In aqueous applications, the discharged phase Zn$_{1.5}$CrOPO$_4$  displays a 340 meV/atom decomposition energy into Zn$^{2+}$, P(OH)$_2$ and Cr(OH)$_4$$^-$, indicating that a phosphor hydroxide-rich surface passivation layer may form. Similarly to the first candidate (CoPO$_4$F, mp-25444), this candidate structurally belongs to the tavorite family. As such, it features two inequivalent Cr sites forming CrO$_6$ octahedra, which share corners with two equivalent CrO$_6$ octahedra and four equivalent PO$_4$ tetrahedra.  
Zn intercalation in this polymorph of CrOPO$_4$ can occur in three symmetrically unique sites, labeled A, B, and C in order of decreasing stability.
ApproxNEB calculations for this material confirm a one-dimensional diffusion pathway (\textbf{PathwayI}), where Zn ions hop between adjacent C and A sites, alternating between consecutive, symmetrically unique 866 meV and 958 meV hops. 
Unfortunately, in this specific polymorph, the ApproxNEB results did not highlight the presence of alternative Zn migration pathways that could expand the dimensionality of the diffusion framework.
However, as shown in Figure \ref{fig:772173_pathways} this pathway can, in principle, offshoot into a 676 meV hop between adjacent B and C sites, providing access to otherwise inaccessible metastable sites and forming an expanded 1D diffusion pathway (\textbf{PathwayII}) that is a superset of \textbf{PathwayI}, increasing the theoretical capacity of the material while maintaining the overall 1D dimensionality of the migration pathway.
This result is consistent with the literature: as further expanded in Section \textbf{Evaluating Material Design Rules}, tavorite-based materials have been associated with lower-dimensional migration pathways due to the high-overlap motifs present between their polyhedra.

On the other hand, the intercalated structure of this second polymorph, mp-26924 (also Zn$_{1.5}$CrOPO$_4$), exhibits an energy above hull = 47 meV/atom
and 300 meV/atom in aqueous media, yielding Cr(OH)$_4$$^-$ and P(OH)$_2$$^-$ as decomposition products.\cite{singh2017, ong2008}
It crystallizes in an orthorhombic symmetry (Pnma space group), with lattice parameters of a = 6.09 \AA, b = 7.16 \AA, c = 8.13 \AA{} $ $ and relative angles of $\alpha$ = $\beta$ = $\gamma$ = 90.00\degree. 
In this structure, Cr$^{5+}$ ions form CrO$_5$ trigonal bipyramids, corner-sharing with four equivalent PO$_4$ tetrahedra. ApproxNEB calculations for this structure in the dilute limit confirm a one-dimensional pathway (\textbf{PathwayI}) defined by four symmetrically equivalent 744 meV hops, where Zn ions migrate between B and C sites, creating a `tunnel' hop between non interconnecting CrO$_5$ polyhedra. There is, however, an alternative 1D pathway (\textbf{PathwayII}) which enables another channel for Zn ion mobility in the structure. This pathway involves two symmetrically inequivalent hops:  a 34 meV hop between A and C sites, and a 951 meV hop between adjacent C sites. This alternative pathway maintains the same overall `inter-tunnel' direction for Zn migration as the primary 744 meV pathway (Figure \ref{fig:26924_pathways}).

Even though the mp-26924 polymorph of CrOPO$_4$ exhibits a migration pathway with a barrier of 774 meV (PathwayI), both polymorphs display similar ApproxNEB migration barriers (951 meV for mp-26924, and 958 meV for mp-772173) for the other migration Pathways (respectively PathwayII for mp-26924 and PathwayI/II for mp-772173), and are classified as 1D diffusers. This result may pose challenges for the performance of the material unless nanosized.\cite{liang2015} 
However, the presence of multiple migration pathways in \textbf{mp-26924} suggests that symmetry breaking within structures with high overlap motifs, such as tavorites, could introduce alternative, potentially lower-energy migration pathways. 

\subsubsection{Mn$_2$(PO$_4$)$_3$ (mp-26062) \label{sec:mn2po43}}
Table \ref{tab:tab1} displays the theoretical properties calculated for NASICON Mn$_2$(PO$_4$)$_3$ (mp-26062). The discharged structure Zn$_{0.5}$Mn$_2$(PO$_4$)$_3$ exhibits an average intercalation voltage of 1.77 V vs Zn/Zn$^{2+}$, a volume change of 20\%, and an aqueous instability = 450 meV/atom against P(OH)$_2$$^-$, Zn$^{2+}$, MnO$_4$$^-$. Its intercalated phase exhibits an energy above the hull 60 meV/atom and a high theoretical gravimetric capacity of 240.02 mAh/g and energy density of 424.23 Wh/kg. 

Mn$_2$(PO$_4$)$_3$ belongs to the trigonal R$\overline{3}$ space group. Consistently with its trigonal symmetry, its calculated lattice parameters a = b = 8.16 \AA, c = 22.25 \AA, and relative angles of $\alpha$ = $\beta$ = 90.00\degree, $\gamma$ = 120.00\degree. Mn$_2$(PO$_4$)$_3$ adopts a 3D structural framework, presenting two inequivalent Mn sites forming MnO$_6$ octahedra, which share oxygen corners with six equivalent PO$_4$ tetrahedra. In this structure, Zn can be intercalated in one of two non-symmetrically equivalent Zn sites, labeled A and B, with the former being lower in energy due to its location in a low-energy cavity (1a Wyckoff position). The ApproxNEB calculations for this material reveal multiple migration pathways, each presenting three hops: two symmetrically equivalent 886 meV hops connecting A and B sites, and one symmetrically unique 894 meV hop between adjacent B sites. Multiple pathways are identified in the host structure, resulting in a 3D migration network (Figure \ref{fig:26062_pathways}). This result is consistent with previous findings for LISICON and NASICON-type frameworks, which demonstrate remarkable Li$^+$ diffusivity and reversibility in repeated cycling due to the topological patterns present in this family of materials, providing high-volume cavities for ion intercalation and migration\cite{masquelier1998, masquelier2013, zhao2018, zhang2019}.

\section{Evaluating Material Design Rules \label{sec:rules}}
In Section \textbf{Introduction}, we highlighted the importance of identifying key features that influence the mobility of the working ion, which are essential for designing performant positive electrode materials. These factors may include structural aspects - such as polyhedral distortions\cite{gao2018, patoux2003}, the density of the host framework\cite{ding2021} (e.g., the volume per anion/cation ratio), and low-overlap motifs that facilitate ion migration\cite{kim2024_1}, as well as more chemical/environmental factors, like the coordination environment and stability of mobile ions\cite{rong2015, van2013, xu2010}, and electrostatic interactions of the host framework~\cite{lu2021, urban2016, kang2006} with the working ion.
The following section will highlight the descriptors that have been identified as potential material design principles in the most promising candidate materials, evaluating their impact on Zn-ion diffusion and the overall performance of candidate cathode materials in ZIBs.

\subsection{Structural and Chemistry Effects.\label{sec:symm}}
As detailed in Section \textbf{Screening Results}, the potential of the screening endeavor in identifying the most promising candidates is confirmed by their similarity to previously investigated structures (e.g. polymorphs) and ICSD prototypes, many of which are recognized as effective materials for intercalation-based energy storage.
In response to the focus on high-voltage, high-energy applications, it is not surprising that all four candidates identified through this screening are polyanion compounds, specifically (fluoro)phosphates. As discussed in Section \textbf{Screening Results}, this trend arises from the strong inductive effect of polyanion groups, where the robust P–O covalency stabilizes the reduced state and enhances the material's redox potential. The presence of fluorine, with its high electronegativity, further amplifies this effect, leading to even higher operating voltages. This trend has been previously observed in both Li-ion \cite{hautier2011, li2017} and Na-ion \cite{ni2017} batteries, where (fluoro)phosphates cathode materials typically exhibit high voltages.

A similar trend is observed for the redox centers: the best-performing candidates present Co$^{4+}$, Cr$^{5+}$, and Mn$^{4+}/$Mn$^{5+}$ as redox centers in their charged state. These ions align with the expected trends for high-voltage cathodes: \cite{liu2016_1, li2024_1} Co$^{4+}$, Cr$^{5+}$, and Mn$^{4+}$/Mn$^{5+}$ are all `late' period IV ions at high oxidation states, which exhibit small ionic radius, high ionization energies and, as a consequence, high redox potential. 
Moreover, the top four candidate materials belong to well-established structural families, such as NASICONs and tavorites.\cite{masquelier2013} For instance, CoPO$_4$F (mp-25444) and CrOPO$_4$ (mp-772173) both structurally belong to the tavorite family (general formula M(XO$_4$)Y, where M = Fe, V, Ti, Mn, Co, ..., X = P, S, W, ..., Y = F, O, OH\cite{masquelier2013}).
Tavorite frameworks are known for their high structural tolerance to intercalation, which results in a higher degree of successful topotactic insertion. 
The insertion sites connect through open channels, enhancing intercalation kinetics and in some cases enable multi-channel ionic transport and fast ion migration.\cite{shen2020}
For these reasons, these materials have been extensively studied for their potential in lithium ion batteries.
While similar compounds (e.g. VPO$_4$F and FePO$_4$F, as well as orthorhombic polymorphs of CoPO$_4$F) have shown promise as cathodes in monovalent ion batteries, \cite{fedotov2016, ahsan2024, okada2005, schoiber2016, hadermann2011, perez2023, fedotov2017, kubota2014} the triclinic polymorphs of CoPO$_4$F and CrOPO$_4$ remain experimentally unexplored. 
However, their frameworks match ICSD entries such as SbOPO$_4$ (mp-9750)\cite{piffard1986}, high-temperature polymorphs of NbOPO$_4$, specifically $\beta$-NbOPO$_4$ (mp-542453)\cite{amos2001, leclaire1986, lu2016}, and with VOPO$_4$, which have recently attracted interest as cathode materials for monovalent\cite{chernova2020, ma2022, shen2023} and divalent\cite{sari2024, zhao2023} ion batteries. Unlike its monoclinic counterpart, the orthorhombic CrOPO$_4$ polymorph (mp-26924) does not strictly belong to the tavorite family, as this form features a more distorted phosphate framework.
However, the structure also aligns with the symmetry group of $\beta$-VOPO$_4$ (mp-25265)\cite{he2016, gopal1972}, a known framework for 3D ion migration\cite{ahsan2024, aparicio2018} and versatile intercalation \cite{chernova2020, ma2022, shen2023}, making it reasonable to anticipate a similar behavior from its CrOPO$_4$ polymorph. Notably, $\epsilon$ and $\delta$-VOPO$_4$ were recently identified computationally as viable Mg-ion cathodes, showing computed NEB barriers of respectively 687 and 588 meV.\cite{sari2024, kim2024_1} Experimental studies on the $\epsilon$ polymorph demonstrated intercalation supporting energy densities exceeding 200 Wh/kg for small particle sizes (~100 nm).\cite{rong2016}

Lastly, the electrochemical performance of NASICON materials (general formula M$_2$(XO$_4$)$_3$, where M = Fe, V, Ti, Mn, Co, ... X = P, S, W, ... \cite{masquelier2013}) has also been the object of several previous investigations, highlighting the influence of structural and compositional descriptors - such as polyhedral distortions and disproportionation reactions at the redox-active sites - on the resulting performance and electrical conductivity of these materials\cite{gao2018, patoux2003} and suggesting various synthetic strategies\cite{liu2022, zhou2023} and design guidelines\cite{hautier2011, wang2023} to optimize the overall performance of these materials.
While specific diffusion pathways for Mn$_2$(PO$_4$)$_3$ have yet to be reported, NASICON-type phosphates have been widely investigated as cathode materials for both Li and Na ion batteries\cite{masquelier1998, masquelier2013}.
Mn$_2$(PO$_4$)$_3$, in particular, shows an ICSD subset match to Nb$_2$(PO$_4$)$_3$ (mp-17242)\cite{leclaire1989}, a known anode material in Li- and Na-ion batteries\cite{patra2023} characterized by diagonal diffusion paths, suggesting that a similar diffusion behavior might be expected for Mn$_2$(PO$_4$)$_3$.
A more detailed description of the computational and experimental research on these materials is detailed in Section \ref{sec:rules_SI} the SI of this work.
Overall, the structural parallels displayed by the candidate materials to studied cathode frameworks show that known structural motifs can be applied for the prediction and design of new intercalation hosts, emphasizing the role of structure as a transferable descriptor for ion mobility.

\subsection{Evolving environment. \label{sec:environment}} 
The changing crystalline environment surrounding the working ion throughout a migration event, known as evolving environment\cite{rutt2022}, is another descriptor affecting the mobility of the working ion.
The origin of this descriptor stems from work by Rong et al\cite{rong2015}, which illustrated a correlation between migration barriers and ion topology in host frameworks. In the paper, the authors suggested that higher barriers often occur in environments in which the preferred coordination number of the working ion matches its effective coordination in the host.
This trend has been confirmed in computational screening studies for Mg$^{2+}$ and Ca$^{2+}$\cite{rutt2022, kim2024_1} electrode materials, motivating further exploration into the behavior of Zn$^{2+}$.  Zn$^{2+}$ is known to exhibit a preferred 4- or 6-fold coordination depending on the anion/base strengths of the surrounding ions and on the spatial constraints of the evolving environment.\cite{brown1988}

\begin{figure}[ht!]
    \centering
    \renewcommand{\arraystretch}{1}
    \setlength{\tabcolsep}{5pt}
    \setlength{\arrayrulewidth}{1pt}

    \begin{tabular}{l!{\color{mycolor}\vrule width 1pt} l}
          \begin{subfigure}[t]{0.49\textwidth}  
         \centering
         \includegraphics[width=\linewidth]{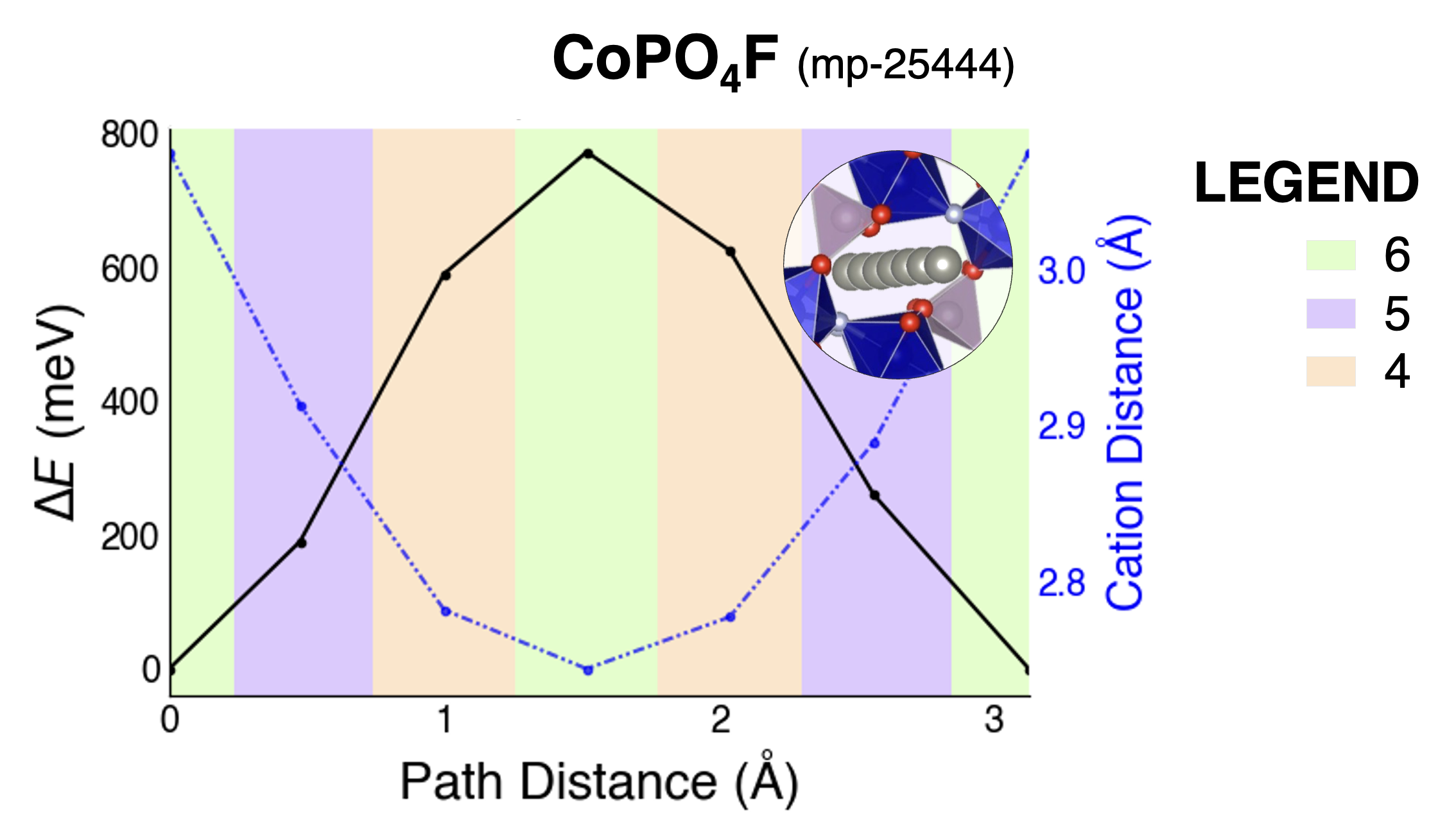}
         \caption{}
         \label{fig:25444_cn}
         \vspace{1em} 
     \end{subfigure}
&
     \begin{subfigure}[t]{0.49\textwidth}
         \centering
         \includegraphics[width=\linewidth]{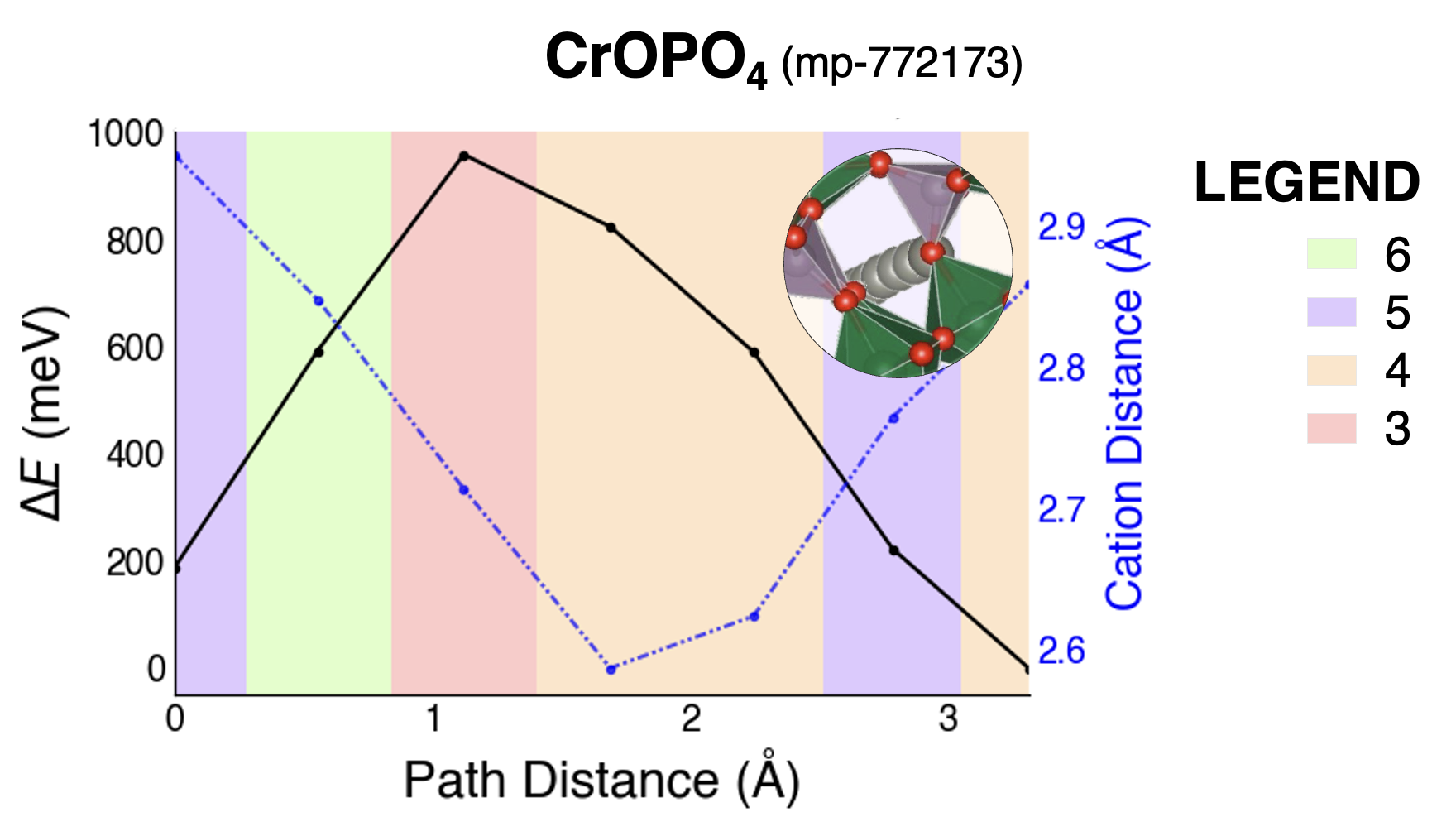}
         \caption{}
         \label{fig:772173_cn}
         \vspace{1em} 
     \end{subfigure}
     \\
     \arrayrulecolor{mycolor}\hline
\\
    \renewcommand{\arraystretch}{1.2}
     \begin{subfigure}[t]{0.49\textwidth}
         \centering
         \includegraphics[width=\linewidth]{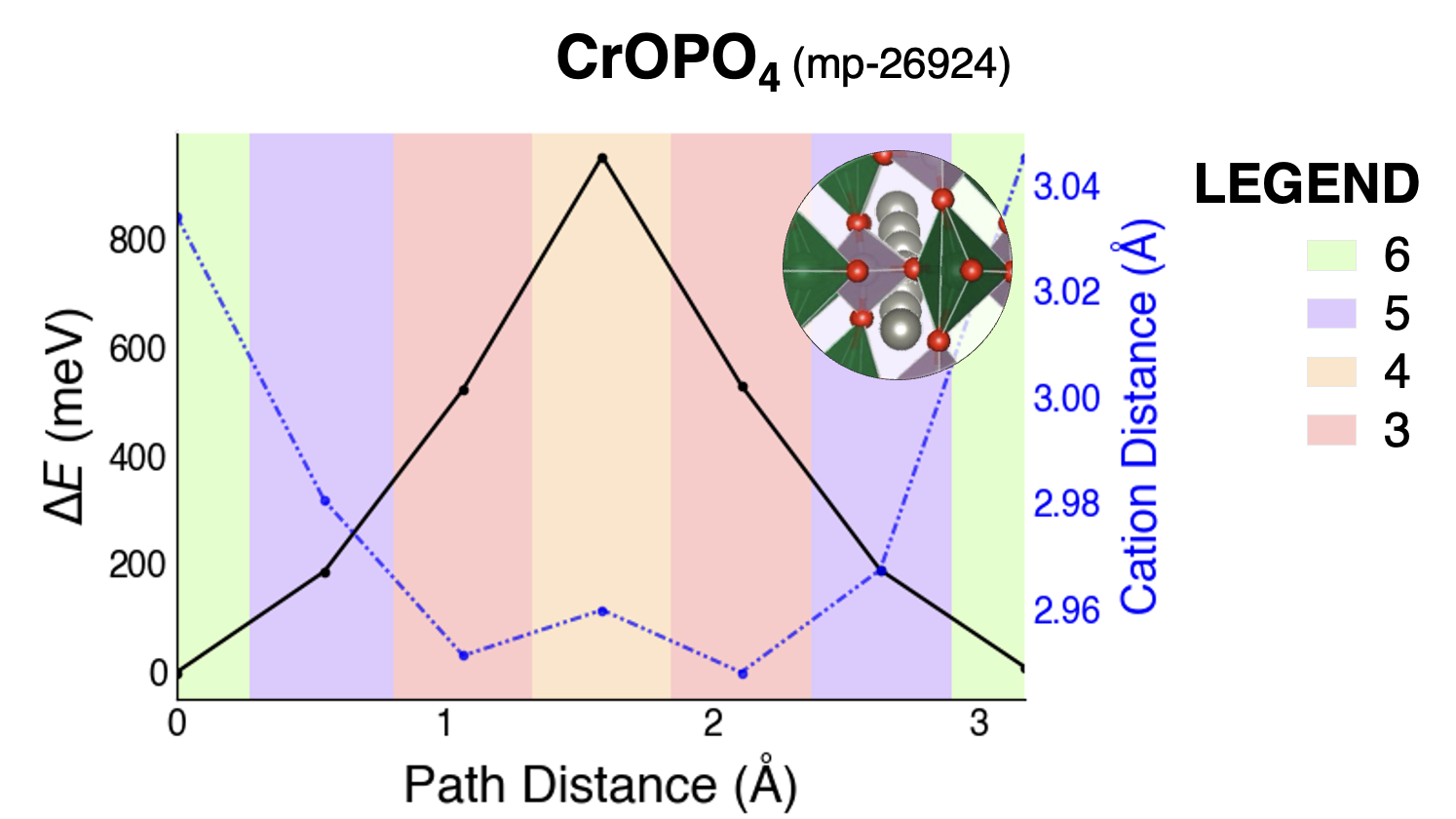}
         \caption{}
         \label{fig:26924_cn}
     \end{subfigure}
&
     \begin{subfigure}[t]{0.49\textwidth}
         \centering
         \includegraphics[width=\linewidth]{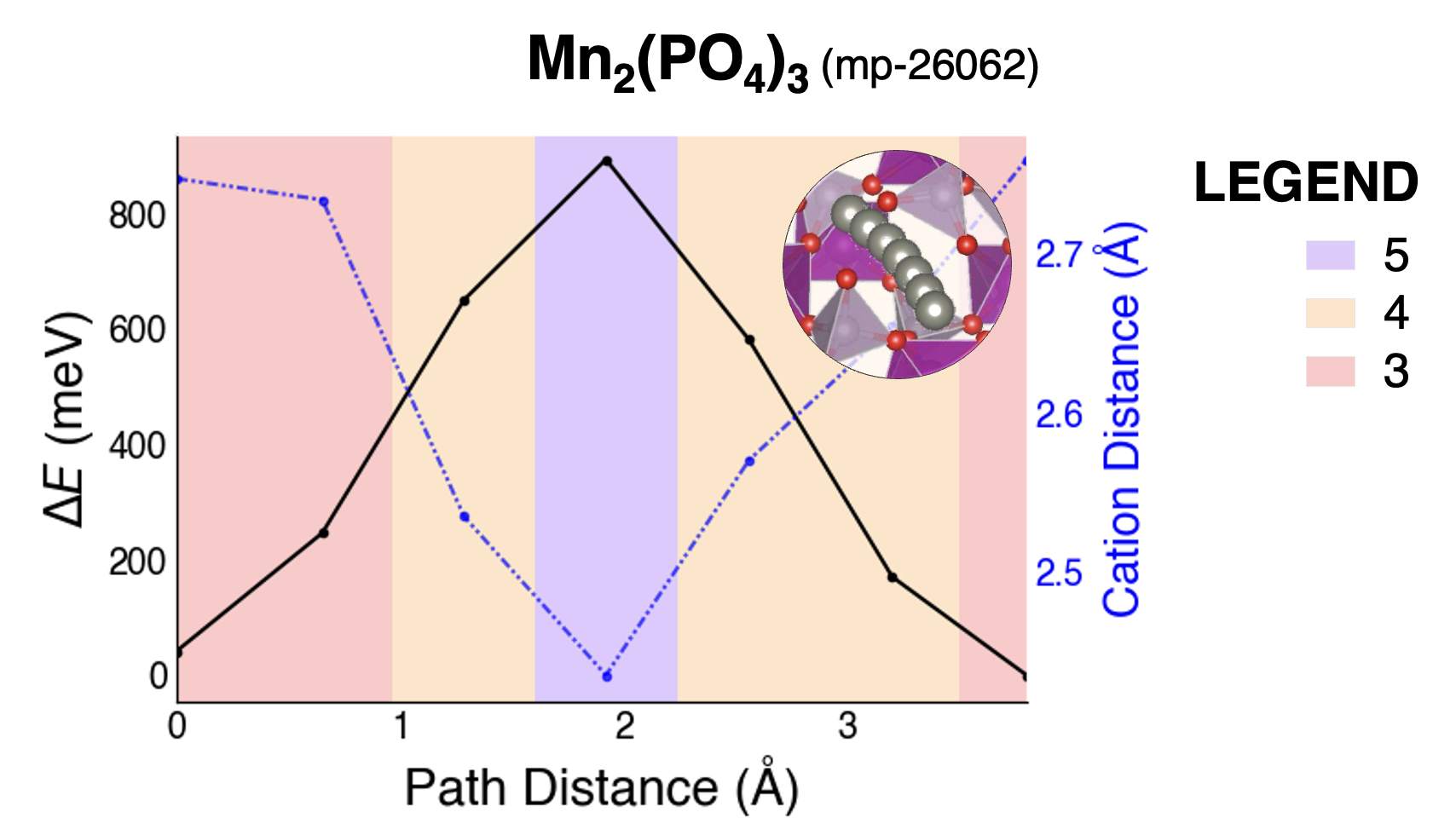}
         \caption{}
         \label{fig:26062_cn}
     \end{subfigure}  
    \end{tabular}
     \caption{Evolving environment and associated Zn$^{2+}$ migration energy as a function of the pathway coordinate in the four best performing candidates: (\ref{fig:25444_cn}) CoPO$_4$F, (\ref{fig:772173_cn}-\ref{fig:26924_cn}) the two polymorphs of CrOPO$_4$, (\ref{fig:26062_cn}) Mn$_2$(PO$_4$)$_3$. Graph of the ApproxNEB energy barrier in the rate limiting step of the migration path (black line with circles), as a function of (i) Zn$^{2+}$ coordination (colored graph area), (ii) Zn$^{2+}$ distance from nearest cation in the host framework (blue dotted line with circles).
     }
     \label{fig:cn_zn}
\end{figure}

In our calculations, we quantified the influence of the evolving environment on the energy landscape of Zn$^{2+}$ migration in candidate materials. In particular, in Figures \ref{fig:25444_cn}-\ref{fig:26062_cn} we show the Zn$^{2+}$ energy landscape of the four best performing candidates in the most kinetically hindered hops (or bottleneck hops), represented using the VESTA software\cite{momma2011}) of the ApproxNEB migration pathways analyzed in Section \textbf{ApproxNEB results}, as these provide an indications of the factors that affect ion mobility in Zn$^{2+}$ diffusion. The plots are color-coded to depict the varying Zn$^{2+}$ coordination at different positions along the migration hops, as analyzed using the \texttt{CrystalNN} algorithm in \texttt{pymatgen}.\cite{ONG2013314, pan2021} They also depict the distance between the mobile Zn$^{2+}$ and the nearest cation (by average) in the host structure.

In general, poor migration is expected to arise in materials where Zn$^{2+}$ resides in strong potential well, leading to restricted mobility. Conversely, improved mobility is expected for topologies where the mobile ion is stabilized by its preferred coordination at the activated site. However, notably, we find that in these materials the strongest influence on the activation barrier is the repulsion between the Zn2+ and the closest cation. 
For instance, in CoPO$_4$F (mp-25444), the highest-energy point occurs when six-fold coordinated Zn squeezes through an anion cavity formed by four O atoms and two F atoms in equivalent, non corner-sharing CoO$_4$F$_2$ octahedra.  In the two CrOPO$_4$ polymorphs and in Mn$_2$(PO$_4$)$_3$, the migration energy bottleneck occurs when Zn migrates through anion cavities formed between redox-active TM polyhedra and PO$_4$ tetrahedra - respectively three O atoms in adjacent CrO$_6$ octahedra in CrOPO$_4$ (mp-772173), resulting in three-fold coordination, four O atoms in CrO$_5$ trigonal bipyramids in CrOPO$_4$ (mp-26924), displaying four-fold coordination, and five O atoms in non-adjacent PO$_4$ tetrahedra in Mn$_2$(PO$_4$)$_3$ (mp-26062), corresponding to five-fold coordination for Zn. As shown in Fig. \ref{fig:cn_zn}, in the majority of cases, the highest-energy points correspond to the closest distance between Zn and its nearest cation and/or to the lowest Zn coordination in the host. Interestingly, at shorter Zn-cation distances ($<$ 2.7 \AA), we observe a modulation between coordination and electrostatics. This modulation is evident in the case of CrOPO$_4$, for which the activation barrier is set for the 3-fold coordinated site, slightly offset from the closest cation-cation distance. The high energy results from a combination of structural constraints, which restrict the available space for Zn diffusion within the host structure, and electrostatic repulsion between Zn and the closest cation. 
Similar effects are well-known in the layered and rocksalt Li-ion cathode materials\cite{rong2015}: first shown by Van der Ven and Ceder \cite{van1999, van2001, van2013}, lithium migration follows a divalent mechanism via a tetrahedral activated state, where the migrating Li$^+$ remains in close proximity to a transition metal. This model predicts maximal lithium diffusivity in partially delithiated phases, aligning with experimental observations. Higher Li diffusion was attributed to Li vacancies, which lower the energy of the activated states by decreasing the electrostatic repulsion between the activated lithium ion and its neighboring cations, and was later confirmed for a variety of layered and spinel materials\cite{urban2016, kang2006, kang2006_1}. 
Additionally, in a similar fashion to Li, we note that at high charge, the repulsive interaction between Zn and the transition metal in the host structure is significantly influenced by the oxidation state of the transition metal, as higher oxidation states will result in higher electrostatic interactions with the migrating ion and, leading to higher activation barriers.\cite{kang2006, van1999, van2001, jang2001}
Interestingly, while current Li-ion cathodes are synthesized in the discharged state—where repulsive interactions intensify as the transition metal valence state increases—the candidate Zn-ion cathodes, for which the computed barriers correspond to the charged state, are expected to exhibit a decrease in cation repulsion with increased intercalation and with the consequent reduction of the redox-active transition metal centers.
Overall, our work underscores the crucial influence of both structural and electrostatic interactions on ion mobility, offering insights for the design of next-generation multivalent battery materials. Indeed, previous studies conducted on Mg$^{2+}$\cite{rutt2022} indicated that the coordination number, while important, might not be an adequate descriptor for the energy profile of the migration pathway, suggesting that the electrostatic landscape of the host material plays a significant role in defining the energetic landscapes (and penalties) of ionic migration.

\section{Conclusions\label{sec:conclusion}}

In this study, we employed a high-throughput computational screening pipeline to identify high performance cathode materials for Zn-ion batteries (ZIBs). Our automated discovery pipeline narrowed down the initial dataset of 163,109 candidate materials from the Materials Project database through increasingly selective and more resource-intensive tiers. The first stage applied property screening based on composition, stability and synthesizability, and practical considerations, focusing on key performance descriptors such as operating voltage windows and gravimetric energy densities. The screening criteria were then integrated with density functional theory (DFT) calculations to evaluate Zn-ion intercalation and diffusion in the materials, further refining the selection through comparison with similar, experimentally synthesized structures from the Inorganic crystal structure database (ICSD).

Focusing on high-voltage cathode applications, the initial dataset was narrowed to four top-performing candidates: CoPO$_4$F (mp-25444), two polymorphs of CrOPO$_4$ (mp-772173, mp-26924), and Mn$_2$(PO$_4$)$_3$ (mp-26062).
Our analysis of these materials for material design descriptors emphasized a correlation between ion mobility and common structural motifs in the host structures: the high voltages presented by the four best candidates are attributed to the stabilizing effect of phosphates and fluorophosphates on high-voltage redox centers through inductive effect, as well as the presence of `late' period IV ions at high oxidation states (Co$^{4+}$, Cr$^{5+}$, and Mn$^{4+}$/Mn$^{5+}$). These trends, previously observed in Li- and Na-ion batteries,\cite{liu2016_1, li2024_1,hautier2011, li2017, ni2017}, are here extended to ZIBs. Furthermore, the presence of feasible intercalation sites/pathways and favorable migration barriers ($<$ 1 eV) are ascribed to structural features of the host frameworks, which present high tolerance to intercalation, fully connected migration networks and deviations from the classical frameworks that introduce distorted morphologies. We find that the repulsion between Zn and the nearest cation correlates the most strongly with the activation energy barrier, with slight modulation for the local environment. This suggests that the anticipated migration barriers may be ameliorated at partial discharge. 

We note that removing materials with multiple movable ions (e.g. Li$^{+}$, Na$^{+}$, Mg$^{2+}$, Zn$^{2+}$, etc) in the pipeline also removes the possibilty of designing for coordinated motion. Future studies may explore alternative screening criteria, including frameworks containing working ions and/or that have demonstrated high performance with other multivalent ions. However, our results emphasize the effectiveness of a structure-based approach for identifying candidate materials, reinforcing the role of framework topology and connectivity in optimizing ion migration through the host structures.

Future investigations on prospective candidate materials should further investigate Zn-ion mobility and kinetics in the identified candidates through more accurate computational investigations, such as climbing-image nudged elastic band (CI-NEB) calculations and \textit{ab initio} molecular dynamics (AIMD). Experimental validations (synthesis, electrochemical characterization) will also be critical to confirm feasibility, with a particular focus on their stability under prolonged cycling.

Overall, this study refined the pre-existent discovery pipeline for cathode materials by linking their performance and ion diffusivity to structural and chemical design principles. By expanding the pipeline to ZIBs, this study has further extended the chemical search space for divalent-ion battery materials, driving the development of safer, cost-effective, and sustainable energy storage technologies. 

\begin{acknowledgement}
This work was intellectually led by the BEACONS Center at UT Dallas supported by the U.S. Department of Defense's Office of Industrial Base Policy and its Manufacturing Capability Expansion and investment Prioritization (MCEIP) office funding of the Batteries and Energy to Advance Commercialization and National Security (BEACONS) at the University of Texas at Dallas. Partial support was obtained from the Energy Storage Research Alliance "ESRA" (DE-AC02-06CH11357), an Energy Innovation Hub funded by the U.S. Department of Energy, Office of Science, Basic Energy Sciences.  Data and workflow algorithms provided and further developed with the aid of the Materials Project, which is funded by the U.S. Department of Energy, Office of Science, Office of Basic Energy Sciences, Materials Sciences and Engineering Division, under Contract No. DE-AC02-05-CH11231: Materials Project Program KC23MP. Our research used resources of the National Energy Research Scientific Computing Center (NERSC), a Department of Energy Office of Science User Facility using NERSC award DOE-ERCAP0026371.

\end{acknowledgement}

\section{Supporting Information}

The Supporting Information is available free of charge at https://pubs.acs.org/doi/xxxx.

Screening criteria employed in the different Tiers; literature review on best candidates (for most common working ions): polymorphs, different compositions, ICSD matches; methodology/computational details employed in the calculations (PDF).

\section{AUTHOR INFORMATION}
\subsection{Corresponding Author}

\textbf{Kristin A. Persson} - Department of Materials Science and Engineering, University of California, Berkeley, California 94704, United States;  Materials Sciences Division, Lawrence Berkeley National Laboratory, Berkeley, California 94720, United States

\subsection{Authors}

\textbf{Roberta Pascazio} - Department of Materials Science and Engineering, University of California, Berkeley, California 94704, United States; Materials Sciences Division, Lawrence Berkeley National Laboratory, Berkeley, California 94720, United States

\textbf{Qian Chen} - Lawrence Berkeley National Laboratory, Berkeley, California 94720, United States, Present address: Toyota Research Institute of North America, Ann Arbor, Michigan 48105, United States

\textbf{Haoming Howard Li} - Department of Materials Science and Engineering, University of California, Berkeley, California 94704, United States; Materials Sciences Division, Lawrence Berkeley National Laboratory, Berkeley, California 94720, United States

\textbf{Aaron D. Kaplan} - Materials Sciences Division, Lawrence Berkeley National Laboratory, Berkeley, California 94720, United States

\subsection{Author Contributions}

The manuscript was written through contributions of all authors. 
R. P. completed the diffusion calculations, data analysis for stability and electrochemical properties and descriptor analysis, visualization, drafting, reviewing and editing of the paper. 
Q.C. designed and performed the screening procedure, the insertion and the preliminary diffusion calculations, data analysis for stability and electrochemical properties, and supervised the preparation and editing of the manuscript.
H.H.L. and A.D.K. carried out the supervision of the insertion and diffusion algorithms and supervised the preparation and editing.
K. A. P. provided supervising and funding at all stages and editing of the manuscript. \\
All authors have given approval to the final version of the manuscript. \\
R.P. and Q.C. contributed equally to this work.

\section{Notes}

The authors do not declare competing financial interest(s) in this manuscript.

\section{ABBREVIATIONS}

ZIB, Zn-ion battery; LIB, Li-ion battery; DFT, density functional theory; PBA, Prussian blue analogues; EDX, energy-dispersive X-ray; MP, Materials Project; TM, Transition metal; NASICON, sodium super-ionic conductors; ICSD, Inorganic Crystal Structure Database; (CI)-NEB; (Climbing-image)-nudged elastic band; AIMD, \textit{ab initio} molecular dynamics; OER, oxygen evolution reaction; SHE, standard hydrogen electrode.

\bibliography{references}

\clearpage

\input{Supporting_Info}

\end{document}

%% file: Supporting_Info.tex
\setcounter{section}{0}
\setcounter{table}{1}
\setcounter{figure}{1}
\setcounter{page}{1}
\renewcommand{\thepage}{S\arabic{page}}
\renewcommand{\thesection}{S\arabic{section}}
\renewcommand{\theequation}{S\arabic{equation}}
\renewcommand{\thetable}{S\arabic{table}}
\renewcommand{\thefigure}{S\arabic{figure}}

\section*{Supporting Information: \\
\Large Towards High-Voltage Cathodes for Zinc-Ion Batteries: Discovery Pipeline and Material Design Rules}

\section{Screening criteria}
The following sections detail the criteria applied for all tiers in the screening protocol.
\subsection{Tier 1: Property Screening \label{sec:tier1_SI}}

\textbf{Composition screening.} The composition screening evaluated the following criteria: (i) the presence of at least one transition metal capable of undergoing reduction reactions upon Zn insertion (Ti, V, Cr, Mn, Fe, Co, Ni, Cu, Nb, Mo, Ru, Ag, W, Re, Sb, and Bi); (ii) the presence of either O or S in the host structure; (iii) the absence of radioactive, toxic, and/or expensive elements, including As, Au, Ir, Pt, Pd, Rh, Tc, and elements with atomic number greater than 83; (iv) the absence of Zn or other intercalating ions, such as H, Li, Na, Ca, Mg, and K, to simplify the evaluation of ion mobility and diffusivity. 
This screening narrowed the pool of potential candidates from 154,718 to 22,769 structures. \\
\textbf{Stability/synthesizability screening.} In this tier, materials were screened for (i) an energy above the hull\footnote{The energy above the hull is defined as the distance of a phase from the hull composed by its most stable phases, and it is a measure of its thermodynamic stability \cite{ong2008}.} $<$ 0.1 eV/atom, in agreement with established thresholds\cite{hautier2012,sun2016}: structures that are unstable in their charged state are unlikely to be stable during the discharge process; (ii) aqueous instability $<$ 0.5 eV/atom, measured at 1.5 V vs. Zn/Zn$^{2+}$ and a pH range of 5-5.5, ensuring that the material does not decompose in water and that stable solid phases are present in decomposed products. The aqueous stability is calculated using the \texttt{PourbaixDiagram} module in \texttt{pymatgen}\cite{ONG2013314,persson2012}. The choice of looser criteria on aqueous stability is motivated by the possible formation of passivation layers on the materials, which would allow for their stability in aqueous media.\cite{singh2017}
This screening reduced the number of candidates from 22,769 to 6,980 viable materials. \\
\textbf{Practical screening.} This stage screened the following properties: (i) a price per unit capacity $<$ \$0.005/mAh, excluding systems with a large fraction of expensive elements from the database; (ii) gravimetric capacity $>$ 100 mAh/g, motivated by the search for high-performance materials.
This final screening tier reduced the optimal candidates from 6,980 to 4,297 structures.
Two further screening criteria were applied to the selected structures: a (iii) structural screening, in which only one representative structure was selected for materials with the same formula/structural framework but different transition metal (TM) ordering or ratios (e.g., NASICON structures (TM)$_2$(PO$_4$)$_3$ with different TM combinations). This screening was performed using structure matching capabilities in \texttt{pymatgen}\cite{ONG2013314}; a (iv) TM screening, in which systems containing Mn, Co, Cr, and Ni were prioritized due to their favorable voltages and high material availability. \\
The 4,297 structures were thus reduced to 1,181 distinct crystal structure types.

Overall, this final screening tier reduced the optimal candidates from 6,980 to 1,181 structures with distinct crystal structures.
\subsection{Tier 2: ion insertion}\label{sec:tier2_SI}
\subsubsection{Additional Screening and Prototype Matches \label{sec:structure_match_SI}}
The 313 intercalated materials presenting successful insertion electrode calculations were further screened based on the same stability and synthesizability criteria employed in \textbf{Tier 1} of the screening, with the addition of the following criteria: (i) an energy above the hull $<$ 0.1 eV/atom for all the structures obtained in the insertion algorithm to construct voltage pairs; (ii) average intercalation voltage $>$ 1.3 V vs. Zn/Zn$^{2+}$; (iii) gravimetric capacity $>$ 160 mAh/g; (iv) energy density $>$ 300 Wh/kg. These criteria were chosen to target high voltage and energy density cathodes.
Finally, we required (v) a volume change upon intercalation lower than 20\% to account for the typically higher volume changes seen in Zn$^{2+}$ intercalation compared to the 5-10\% common in LIBs\cite{arthur2014, sun2016, WANG2017_1, OKADA2001}.
In addition to these requirements, the best candidates were tested for their conversion voltage using data from the Materials Project phase diagram\cite{ong2008,ong2010pd2} and \texttt{pymatgen}\cite{ONG2013314}. The cutoff chosen for the conversion voltage was such that the difference between the conversion and intercalation voltages could not exceed 0.5 V. The threshold of 0.5 V is based on DFT-calculated data which has been benchmarked on experimental values\cite{hannah2018}.

These conditions ensured that the candidate cathodes fit within high performance and stability standards, as well as within the thermodynamic stability window of common aqueous electrolytes\cite{MING201958, singh2017}. More, they introduce new screening criteria which had not been considered in previous multivalent cathode pipelines\cite{rutt2022, kim2024_1}. \\
This screening Tier reduced the 226 candidates obtained in Tier 2 to the 37 best performing materials. 
In particular, of the 37 materials, 33 (89\%) met the conversion voltage criterion. This result represents a significant improvement over Tier 2 candidates (for which only 51\%, corresponding to 116 materials had passed this threshold), underscoring the effectiveness of using thermodynamic criteria to identify promising materials. 

\section{Literature Background \label{sec:rules_SI}}
The following section provides literature background on the state-of-the-art research on the most promising candidate materials and their ICSD matches in batteries involving Zn and other working ions.

As detailed in Section \textbf{Screening Results}, the top four candidate materials belong to well-established cathode frameworks for multivalent ion batteries, such as NASICONs and tavorites\cite{masquelier2013}.

CoPO$_4$F (\textbf{mp-25444}) and CrOPO$_4$ (\textbf{mp-772173}) both structurally belong to the tavorite family (general formula $M$($X$O$_4$)$Y$, where $M$ = Fe, V, Ti, Mn, Co, ..., $X$ = P, S, W, ..., and $Y$ = F, O, OH\cite{masquelier2013}) 
Tavorite frameworks have been extensively studied for their high structural tolerance to intercalation and open channels for ion diffusion, enhancing intercalation kinetics and potentially enabling multidirectional transport.\cite{shen2020}

Notable examples for $Y$ = F which are chemically and structurally similar to CoPO$_4$F include VPO$_4$F and FePO$_4$F, which have shown promise as cathodes in sodium-ion batteries.\cite{fedotov2016, ahsan2024} \\
For instance, Na$_2$FePO$_4$F exhibits a notable experimental discharge capacity of 117 mAh$/$g, good capacity retention, and a small volume change during the charge/discharge process ($\sim$4\%),\cite{ahsan2024} while vanadium-based polyanion compounds of general formula $A$VPO$_4$F ($A$ = Li, K) have been synthesized using a microwave-assisted solvothermal approach and explored for their electrochemical behavior in Li and Na-ion batteries, maintaining 75\% of the initial specific capacity (111 mAh$/$g) up to 100 cycles.\cite{fedotov2016}
Building on the success of VPO$_4$F and FePO$_4$F, several orthorhombic polymorphs of CoPO$_4$F (Pnma, Pbcn) have been synthesized through solid-state reactions and investigated as high-voltage cathodes for monovalent ions.\cite{okada2005, schoiber2016, hadermann2011, perez2023, fedotov2017, kubota2014}. 
The Pnma phase, in particular, has undergone extensive chemical and thermal stability testing and has a theoretical capacity of 287 mAh/g for two Li$^+$ ions at potentials between 4.2–5.1 V vs. Li/Li$^+$, achieving an experimental capacity of 140 mAh/g for one Li atom by mitigating the influence of side reactions.\cite{perez2023, fedotov2017}
Slightly lower theoretical (122 mAh$/$g per Na atom, or 244 mAh$/$g for 2 Na) and experimental capacities (213 mAh$/$g  at an average potential of 4.3 V vs. Na/Na$^+$) were reported for the sodiated version of this polymorph, Na$_2$CoPO$_4$F\cite{kubota2014}. 
However, these capacities are experimentally limited by the de/intercalation of the second working ion atoms, which fall outside the voltage stability range of most commercial electrolytes.
Improvements have been seen with \textit{in-situ} carbon modification, achieving 107 mAh/g per Na atom\cite{zou2015} in Na$_2$CoPO$_4$F by coating the particles and employing a high-voltage electrolyte.

In a parallel fashion to what is reported for CoPO$_4$F, for CrOPO$_4$, similar structures of the general formula $M$OPO$_4$ - most notably VOPO$_4$ polymorphs - have recently gained prominence as cathode materials for both monovalent\cite{chernova2020, ma2022, shen2023} and divalent\cite{sari2024, zhao2023} ion batteries, showing particular promise as Mg cathodes (for which theoretical capacities of 288 mAh/g have been calculated at an average voltage of 2.7 V vs. Mg/Mg$^{2+}$\cite{sari2024}). 
These successes have inspired further exploration of substituted V$_yM_{1-y}$OPO$_4$ compounds, such as LiV$_y$Cr$_{1-y}$OPO$_4$, with promising results: good capacity retention (92\% after 50 cycles), although parasitic side reactions linked to structural strain have been observed in the system.\cite{kaplan2021}. \\
While the triclinic polymorphs of CoPO$_4$F and CrOPO$_4$ have both been previously explored computationally in high-throughput searches for Li-ion battery cathodes (CoPO$_4$F showed a 7\% volume change and 589 Wh/kg energy density, and was predicted as likely synthesizable - i.e. predicted to release less than 30 meV per atom upon decomposition.
CrOPO$_4$, on the other hand, showed a 6\% volume change),\cite{mueller2011} CoPO$_4$F (\textbf{mp-25444}) and CrOPO$_4$ (\textbf{mp-772173}) have yet to be synthesized or tested for electrochemical performance.

However, their frameworks match ICSD entries such as SbOPO$_4$ (\textbf{icsd-201743}, \textbf{mp-9750}), a high-temperature polymorph of NbOPO$_4$, $\beta$-NbOPO$_4$ (\textbf{icsd-93766}, \textbf{icsd-40870}, \textbf{icsd-93767}, \textbf{icsd-252566}, \textbf{mp-542453}), and the $\alpha$-polymorph of VOPO$_4$, all of which have been previously investigated as intercalation electrodes.\cite{lu2016, sari2024, ahsan2024}
As stated previously, VOPO$_4$ polymorphs ($\alpha$-, $\beta$-, and $\epsilon$-VOPO$_4$ in particular) have been widely investigated as cathodes for Li- and Na-ion batteries\cite{chernova2020, ma2022, shen2023}. For divalent ions, $\epsilon$-VOPO$_4$ shows promise as a Mg cathode, with a theoretical capacity of 288 mAh/g at an average voltage of 2.7 V vs. Mg/Mg$^{2+}$\cite{sari2024}. Similarly, $\alpha$- and $\delta$-VOPO$_4$ have been proposed as Ca cathodes, with $\delta$-VOPO$_4$ also being tested for aqueous Zn-ion batteries, achieving a capacity retention of 91 mAh/g after 1000 cycles at an average voltage of 1.46 V vs. Zn/Zn$^{2+}$\cite{zhao2023}. 
\newline 

Unlike its monoclinic counterpart, the orthorhombic CrOPO$_4$ polymorph (\textbf{mp-26924}) does not strictly belong to the tavorite family, as this form features a more distorted phosphate framework deviating from the classic tavorite arrangement.

However, an ICSD match aligns this structure with the symmetry group of the $\beta$-polymorph of VOPO$_4$, (\textbf{icsd-291605}, \textbf{icsd-9413}, \textbf{mp-25265}), whose properties have been discussed previously. $\beta$-VOPO$_4$ is a known framework for 3D ion migration, showing preferential ion diffusion along the $b$-axis\cite{sari2024, ahsan2024, aparicio2018}, and a widely-employed cathode for multivalent ions (Mg, Ca), as well as monovalent ions like Li and Na, \cite{chernova2020, ma2022, shen2023}, showing capacities of 118.6 mAh/g at average 4 V vs. Li/Li$+$ \cite{ren2009} and making it reasonable to anticipate a similar behavior from the CrOPO$_4$ polymorph.
\newline

Lastly, Mn$_2$(PO$_4$)$_3$ (\textbf{mp-26062}) structurally belongs to NASICON materials (general formula $M_2$($X$O$_4$)$_3$, where $M$ = Fe, V, Ti, Mn, Co, ... $X$ = P, S, W, ... \cite{masquelier2013}).
NASICON materials are known electrode materials, and various synthetic strategies (carbon coating, doping) and theoretical guidelines\cite{hautier2011, wang2023} have been developed to enhance their electrochemical performance\cite{liu2022, zhou2023}, their structural stability and reversibility during cycling, particularly in applications for Li and Na-ion batteries.
While specific diffusion pathways for Mn$_2$(PO$_4$)$_3$ have yet to be reported, NASICON-type phosphates with characteristic rhombohedral symmetry (e.g. Fe$_2$(PO$_4$)$_3$, Ti$_2$(PO$_4$)$_3$) have been widely investigated as cathode materials for both Li and Na ion batteries.
For instance, Li$_3$Fe$_2$(PO$_4$)$_3$ demonstrates a theoretical capacity of 128 mAh/g at an average voltage of 2.85 V vs. Li/Li$+$\cite{masquelier1998, masquelier2013}, while NaTi$_2$(PO$_4$)$_3$ has a theoretical capacity of 133 mAh/g at an average potential of 2.1 V vs. Na/Na$+$\cite{thirupathi2023}. Both materials show excellent reversibility over repeated cycling.\cite{zhao2018, wu2019} 
In addition to these well-studied systems, a number of NASICON-type phosphates incorporating Mn and other transition metals have also been explored: a theoretical capacity of 117 mAh/g at 2.1 V vs. Na/Na$+$ has been reported for Na$_3$MnTi(PO$_4$)$_3$ for the Mn-based redox couples \cite{gao2016}, and a satisfactory theoretical capacity of 111 mAh/g (attributed to the reversible intercalation/deintercalation of two Na-ions), with a high redox potential of 3.6 V vs. Na/Na$+$. \cite{li2018, zhou2016} Good performances were also reported for high-entropy NASICON phosphates of mixed compositions.\cite{wu2022}

Mn$_2$(PO$_4$)$_3$ matches to an ICSD prototype, Nb$_2$(PO$_4$)$_3$ (\textbf{icsd-65658}, \textbf{mp-17242}), a known anode material in Li- and Na-ion batteries\cite{patra2023} characterized by high capacities (227 mAh/g at average voltage of 1.86 V vs. Li/Li$+$ for Li, 179 mAh/g at at average voltage of 1.46 V vs. Na/Na$+$ for Na for the first charge/discharge cycle, with good capacity retention, $\sim$ 60.2\% after 200 cycles) and diagonal diffusion paths, suggesting that a similar diffusion behavior might be expected for Mn$_2$(PO$_4$)$_3$.

\section{Methodology/Computational Details \label{sec:computational_deets_SI}}
All electronic structure calculations were performed using the Perdew-Burke-Ernzerhof (PBE) generalized gradient approximation (GGA) \cite{perdew1996} in the Vienna \textit{ab initio} simulation package (VASP) \cite{kresse1993,kresse1994,kresse1996_1,kresse1996} version 6.3.2.
For the insertion electrode calculations, accurate comparable energies across chemical systems were obtained by employing PBE in the absence of transition metal oxides and sulfides, PBE$+U$ otherwise, and mixing energies from the two calculation methodologies using the established procedure in Materials Project \cite{jain2013,jain2011}.
All ApproxNEB calculations employed PBE without a Hubbard $U$ correction.
Although electronic self-interaction error tends to be highest in stretched radical bonds typical of a transition state \cite{kaplan2023}, the optimal Hubbard $U$ value to eliminate this self-interaction error cannot be calculated easily in high throughput and would require, e.g., a costly calculation via linear response \cite{cococcioni2005}. 
More, using a fixed value of $U$ which is not optimized along the migration path has been shown not to be more performant in predicting accurate barriers via NEB and ApproxNEB \cite{sari2024, morgan2004, liu2015, dathar2011, ong2011}.
Chosen numerical and convergence parameters for these calculations are identical to those in previously published work\cite{kim2024, kim2024_1}, employing the \texttt{MPRelaxSet} as implemented in \texttt{pymatgen} \cite{ONG2013314}. 
The critical electronic self-consistency (\texttt{INCAR}) settings are summarized as follows: structural relaxation was performed until the total energy was converged within $5 \times 10^{-4}$ eV and the forces on each atom were converged within 0.05 eV/\AA{}; the plane-wave energy cutoff was set to 520 eV; all calculations used Gaussian smearing of the Fermi surface with smearing width 0.05 eV.
The $k$-point density was set to be 64 points$\cdot \mathrm{\AA{}}^3$, using the ``reciprocal\_density'' tag in \texttt{pymatgen}.
Consistent with Materials Project calculations and previous ApproxNEB calculations, we used the original PBE pseudopotentials (POTCARs) with release snapshot ``06-05-2010''.
The ApproxNEB workflow implementation in the \texttt{atomate} workflow orchestration package \cite{kiran2017} is described in Ref. \citenum{rutt2022}.